\documentclass[useAMS,usenatbib,usegraphicx,onecolumn]{mn2e}

\usepackage{amssymb,amsfonts,amsmath,amstext,amsgen,amsopn,amsxtra,indentfirst,lscape,times,rotating}

\title[Gliese 581g as a scaled-up Earth]{Gliese 581g as a scaled-up version of Earth: \\ atmospheric circulation simulations}
\author[Heng \& Vogt]{Kevin Heng$^{1}$\thanks{E-mail: kheng@phys.ethz.ch (KH)} and Steven S. Vogt$^{2}$\thanks{E-mail: vogt@ucolick.org (SSV)}\\
$^{1}$Zwicky Fellow, ETH Z\"{u}rich, Institute for Astronomy, Wolfgang-Pauli-Strasse 27, CH-8093, Z\"{u}rich, Switzerland\\
$^{2}$UCO/Lick Observatory, University of California, 1156 High Street, Santa Cruz, CA. 95064, U.S.A.}

\begin{document}

\date{Submitted 2010 October 22.  Re-submitted 2010 December 1, 2011 February 18 and 2011 March 30.  Accepted 2011 April 5.}

\pagerange{\pageref{firstpage}--\pageref{lastpage}} \pubyear{2010}

\maketitle

\label{firstpage}

\begin{abstract}
We use three-dimensional simulations to study the atmospheric circulation on the first Earth-sized exoplanet discovered in the habitable zone of an M star.  We treat Gliese 581g as a scaled-up version of Earth by considering increased values for the exoplanetary radius and surface gravity, while retaining terrestrial values for parameters which are unconstrained by current observations.  We examine the long-term, global temperature and wind maps near the surface of the exoplanet --- the climate.  The specific locations for habitability on Gliese 581g depend on whether the exoplanet is tidally-locked and how fast radiative cooling occurs on a global scale.  Independent of whether the existence of Gliese 581g is confirmed, our study highlights the use of general circulation models to quantify the atmospheric circulation on potentially habitable, Earth-sized exoplanets, which will be the prime targets of exoplanet discovery and characterization campaigns in the next decade.
\end{abstract}


\section{Introduction}

The next frontier in extrasolar planet-hunting is the discovery and characterization of Earth-sized exoplanets --- ``exo-Earths".  A particularly promising route is to search for such planets around nearby M stars.  M dwarf stars have several unique attributes that are driving exoplanet studies and astrobiology, as well as next-generation interferometry and direct imaging missions; they constitute at least 72\% of nearby stars.  As the least massive stars, they have the greatest reflex motion due to an orbiting exoplanet.  Furthermore, the classical habitable (liquid water) zone around M dwarfs is typically located in the range $\sim 0.1$--0.2 AU, corresponding to orbital periods of $\sim 20$ to 50 days --- well-matched to the capabilities of ground-based precision-Doppler surveys.  With such short periods, hundreds of cycles of an exoplanet, with a mass of several times that of Earth, can be obtained within a decade, realizing factors of at least 10 in increased sensitivity for strictly periodic Keplerian signals and enabling Doppler reflex barycentric signals as small as 1 m s$^{-1}$ to be recovered even in the presence of similar-amplitude stellar jitter and Poisson noise.  Although these attributes have only recently become widely recognized by the astronomical community \citep{scalo07,tarter07,char09}, many of the nearest M stars have been prime targets for scrutiny by leading precision-radial-velocity surveys for over a decade now.

One of the most enticing and proximate exoplanetary systems being scrutinized is Gliese 581, with at least four exoplanets \citep{mayor09} orbiting a nearby (6.3 pc) M3V star.  Two of the exoplanets announced by \cite{mayor09} are apparently ``super-Earths" that straddle its habitable zone \citep{sel07}.  Recently, \cite{vogt10} announced two more exoplanet candidates orbiting this star --- one with a minimum mass of $3.1~M_\oplus$ (Gliese 581g) and an orbital distance of about 0.15 AU, placing it squarely within the habitable zone of its parent star.  It is generally accepted \citep{lammer10} that, for stellar masses below $0.6~M_\odot$, an Earth-mass exoplanet orbiting anywhere in the habitable zone becomes tidally locked\footnote{In this paper, we use the terms ``tidal locking" and ``synchronous rotation" synonymously, but this is strictly speaking only correct for an exoplanet on a circular orbit, which is our implicit assumption.} or spin-synchronized within the first Gyr of its origin, such that it keeps one face permanently illuminated with the other in perpetual darkness.  Such tidal locking will greatly influence the climate across the exoplanet and figures prominently in any discussion of its potential habitability.

\begin{figure}
\begin{center}
\includegraphics[width=0.48\columnwidth]{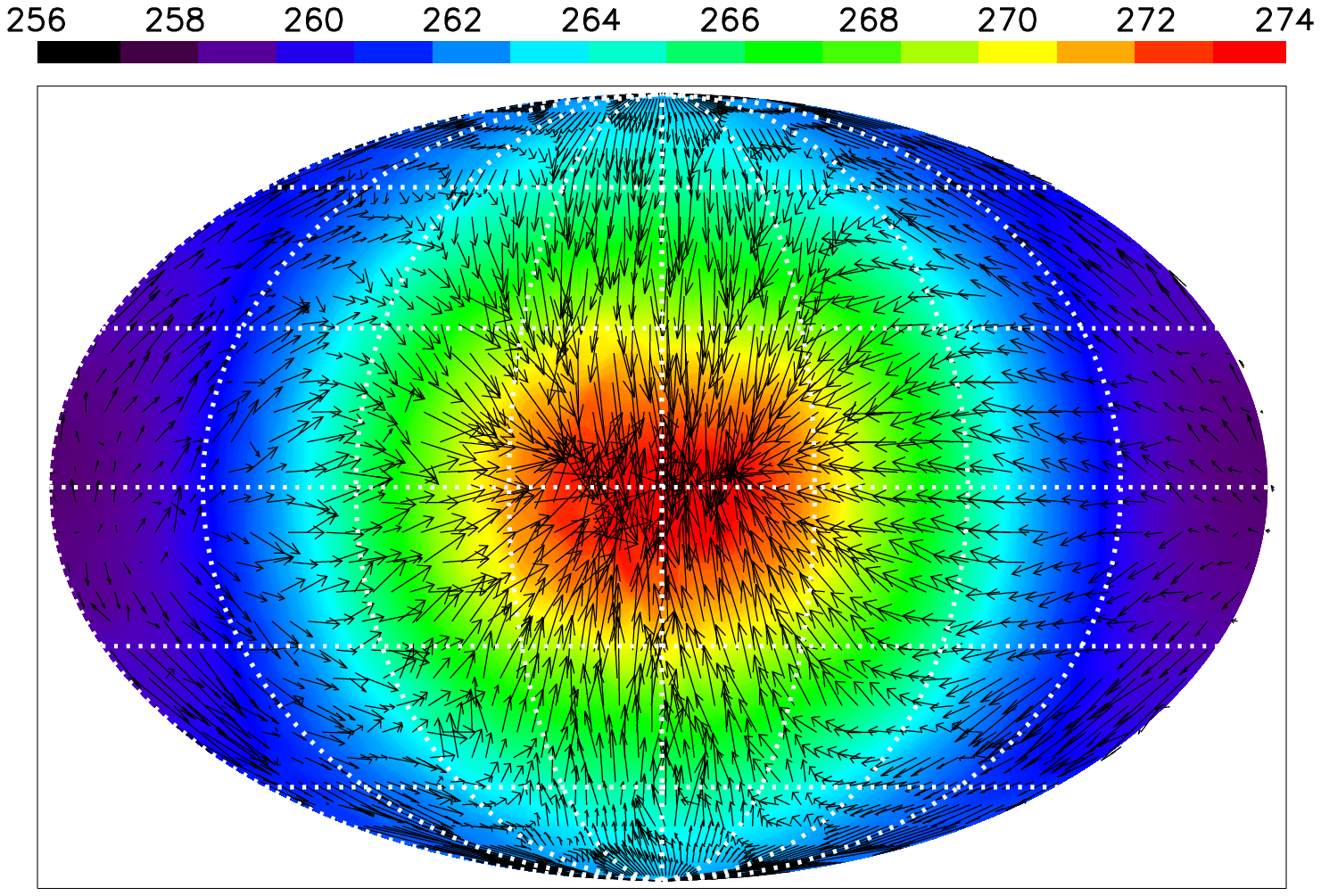}
\includegraphics[width=0.48\columnwidth]{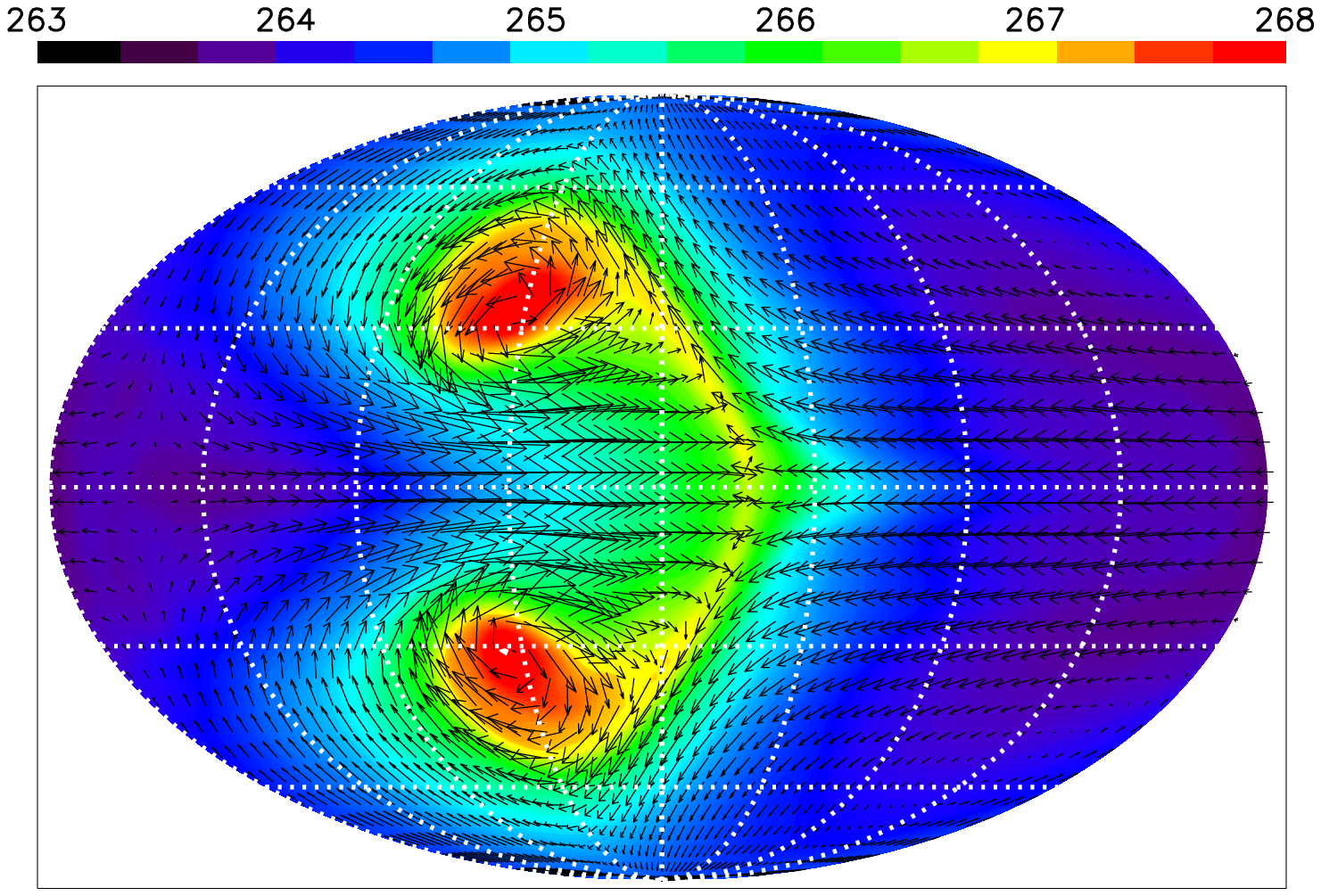}
\end{center}
\vspace{-0.2in}
\caption{Mollweide projection of the temperature and velocity fields near the surface ($P = 0.95$--1 bar) of Gliese 581g as modelled by our atmospheric circulation simulations assuming tidal locking.  Left: simulation with the fiducial value of the radiative cooling time used in the Held-Suarez benchmark for Earth.  Right: simulation with the radiative cooling time lengthened by the ratio of the orbital period of Gliese 581g to the rotational period of Earth (a factor of 36.562).  Colors denote temperature (in K) and the arrows represent the direction of the velocity field.  The resolution adopted is T63L20 ($192 \times 96 \times 20$).  The snapshots are taken at 1200 and 3000 Earth days, for the left and right panels respectively, after the start of the simulations.}
\label{fig:mollweide}
\vspace{-0.1in}
\end{figure}

Independent of whether the existence of Gliese 581g is eventually confirmed, such discoveries motivate the study of atmospheric circulation on exo-Earths using three-dimensional simulations \citep{showman10}, which is the focus of the present paper.  \emph{Our underlying philosophy is to use Gliese 581g as a test bed and explore its atmospheric circulation --- in the absence of observational constraints, we assume parameter values appropriate to the terrestrial atmosphere.}  Unlike in previous work (e.g., \citealt{j97,j03}), we study only the essential dynamics of the atmosphere, choosing not to model the radiative transfer and atmospheric chemistry, an approach which is commensurate with the quality of data currently available for Gliese 581g.  We examine a suite of models both with and without the assumption of tidal locking.  By systematically varying a set of physical parameters, we determine the major and minor parameters involved.  We are primarily interested in the long-term, quasi-stable, large-scale circulation patterns --- the \emph{climate} --- as opposed to the short-term temporal variations (the weather; \citealt{po84}).\footnote{\cite{po84} regard the weather and climate to be initial and boundary value problems, respectively.}  We describe our methods in \S\ref{sect:methods}, present our results in \S\ref{sect:results} and discuss their implications in \S\ref{sect:discussion}.

\section{Methodology}
\label{sect:methods}

\subsection{The GFDL-Princeton \textit{Flexible Modeling System}}

We implement the spectral dynamical core of the \textit{Flexible Modeling System} \texttt{(FMS)} developed by the Geophysical Fluid Dynamics Laboratory at Princeton University \citep{gs82,anderson04,hmp11}.  Dynamical cores are codes that deal with the essential dynamics of atmospheric circulation, treat radiative cooling in a simplified manner (via Newtonian relaxation) and omit atmospheric chemistry \citep{hs94}.  The governing equations solved are the primitive equations of meteorology, where the key assumption made is that of vertical hydrostatic equilibrium \citep{sma63,sma64,wp05,vallis06}.  The boundary-layer friction between the terrestrial atmosphere and surface is treated using a simple linear prescription known as Rayleigh drag/friction.

Following \cite{hs94} and \cite{hmp11}, the first $t_{\rm init}=200$ Earth days of the simulations are discarded (unless otherwise stated); they are then run for $t_{\rm run}=1000$ Earth days.  Our results do not depend on the choice of $t_{\rm init}$ as long as it is chosen to be long enough that transient features due to initialization have been disregarded.  We have verified this statement by performing multiple simulations with larger values of $t_{\rm init}$ and witnessed no difference in our results.  Choosing a larger value of $t_{\rm run}$ allows zonal-mean flow quantities to be averaged over a longer period of time, thus producing smoother profiles as functions of latitude and vertical pressure but otherwise yielding no qualitative difference in the results.

The numerical resolution adopted is T63L20 ($192 \times 96 \times 20$), which corresponds to a horizontal resolution of about 300 km.  By contrast, the vertical pressure scale height is 
\begin{equation}
H = \frac{kT}{m g_p} \approx 60 \mbox{ km} ~\left( \frac{T}{200 \mbox{ K}} \right) \left( \frac{m}{2m_{\rm H}} \frac{g_p}{14.3 \mbox{ m s}^{-2}} \right)^{-1},
\end{equation}
where $k$ is the Boltzmann constant, $T$ is the temperature, $m$ is the mass of the atmospheric molecule(s), $g_p$ is the acceleration due to gravity and $m_{\rm H}$ is the mass of the hydrogen atom.  For comparison, we note that the simulations of \cite{j97} and \cite{j03} use resolutions of T10L10 ($32 \times 16 \times 10$) and T21L22 ($64 \times 32 \times 22$), respectively.  All of the simulations are started from an initial state of windless isothermality ($\vec{v}_{\rm init}=0$, $T_{\rm init}=264$ K) and executed with constant time steps of $\Delta t = 900$ s (i.e., $\sim 10^5$ time steps in total).  The value of $T_{\rm init}$ was again chosen following \cite{hs94} and \cite{hmp11}.  Our conclusions do not depend on this choice as long as quasi-equilibrium is attained.  However, we note that in their simulations of hot Jupiter atmospheres, \cite{tc10} find the resulting structure of the atmospheric circulation to be qualitatively different depending on the assumed initial conditions.  While it is likely that sensitivity to initial conditions scales with the depth of the atmosphere modelled, we consider such an exploration to be beyond the scope of our study.

\begin{figure}
\begin{center}
\includegraphics[width=0.48\columnwidth]{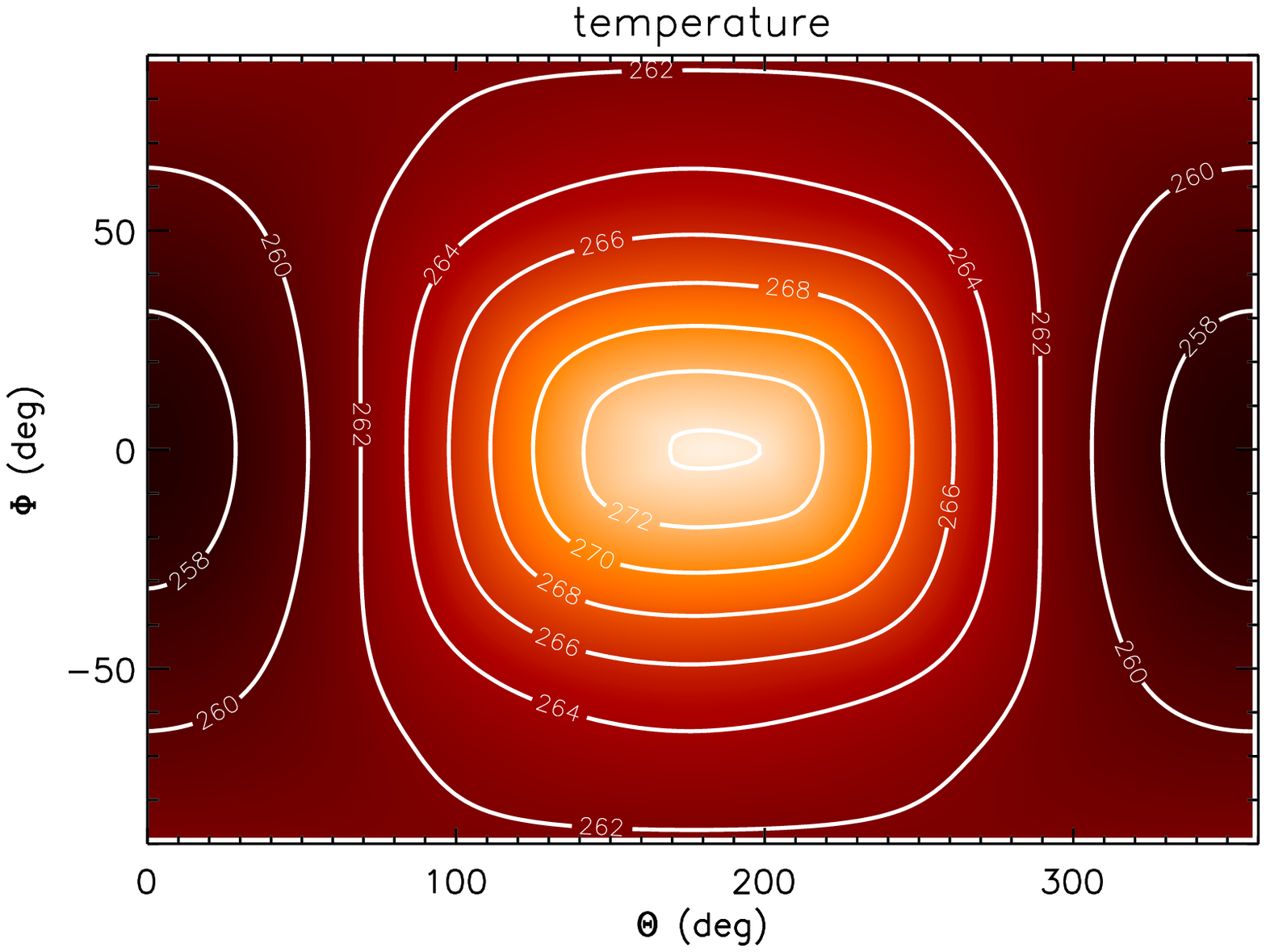}
\includegraphics[width=0.48\columnwidth]{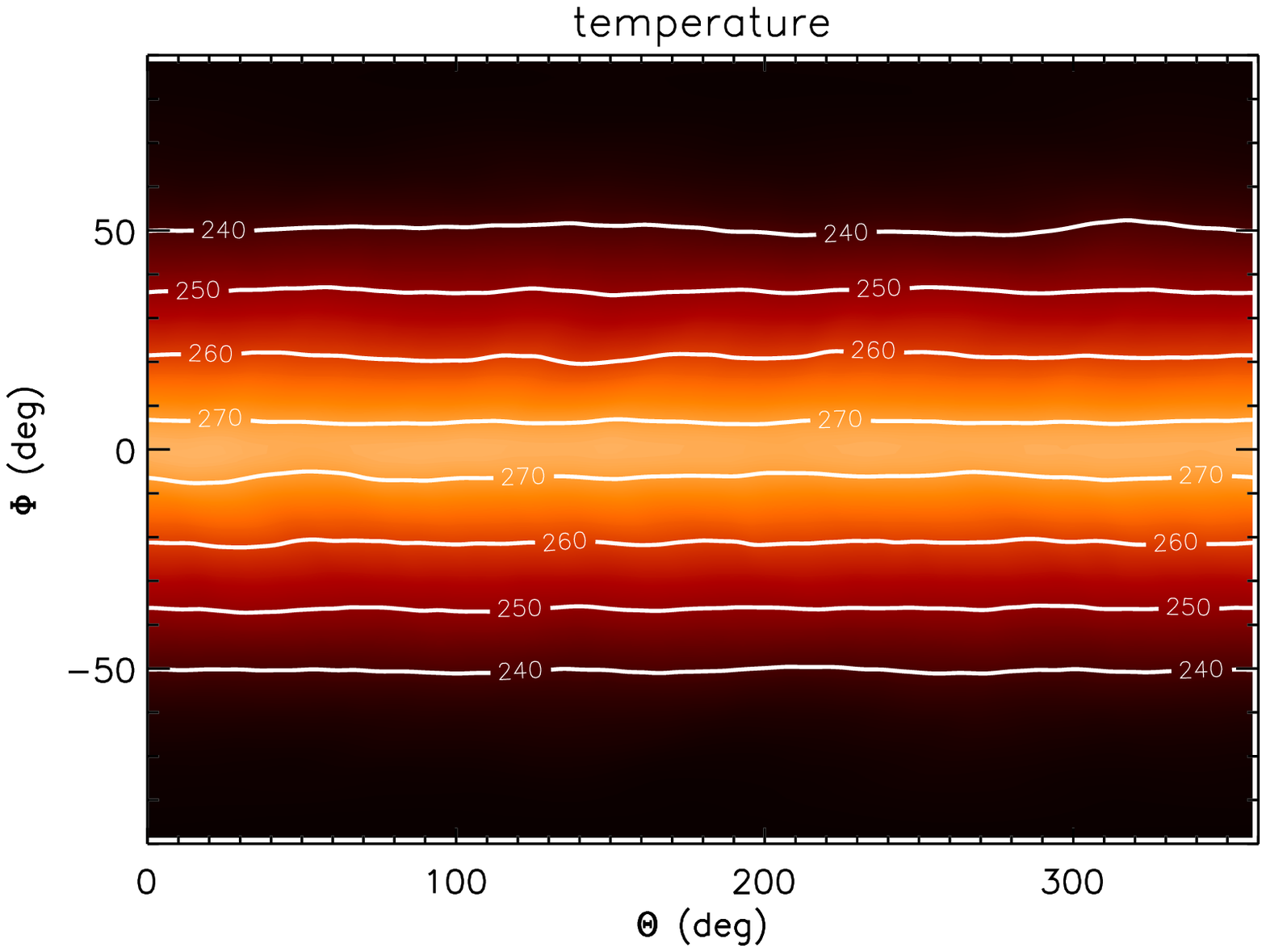}
\end{center}
\vspace{-0.2in}
\caption{Long-term, global temperature maps (in K) near the surface ($P=0.95$--1 bar) of Gliese 581g.  Left: with tidal locking.  Right: a planetary rotation is one Earth day.  The substellar point is located at $\Theta=180^\circ$ and $\Phi=0^\circ$.  These maps are averaged over 1000 Earth days, where the first 200 days of the simulation are discarded.}
\label{fig:temperature}
\vspace{-0.1in}
\end{figure}

\begin{figure}
\begin{center}
\includegraphics[width=0.48\columnwidth]{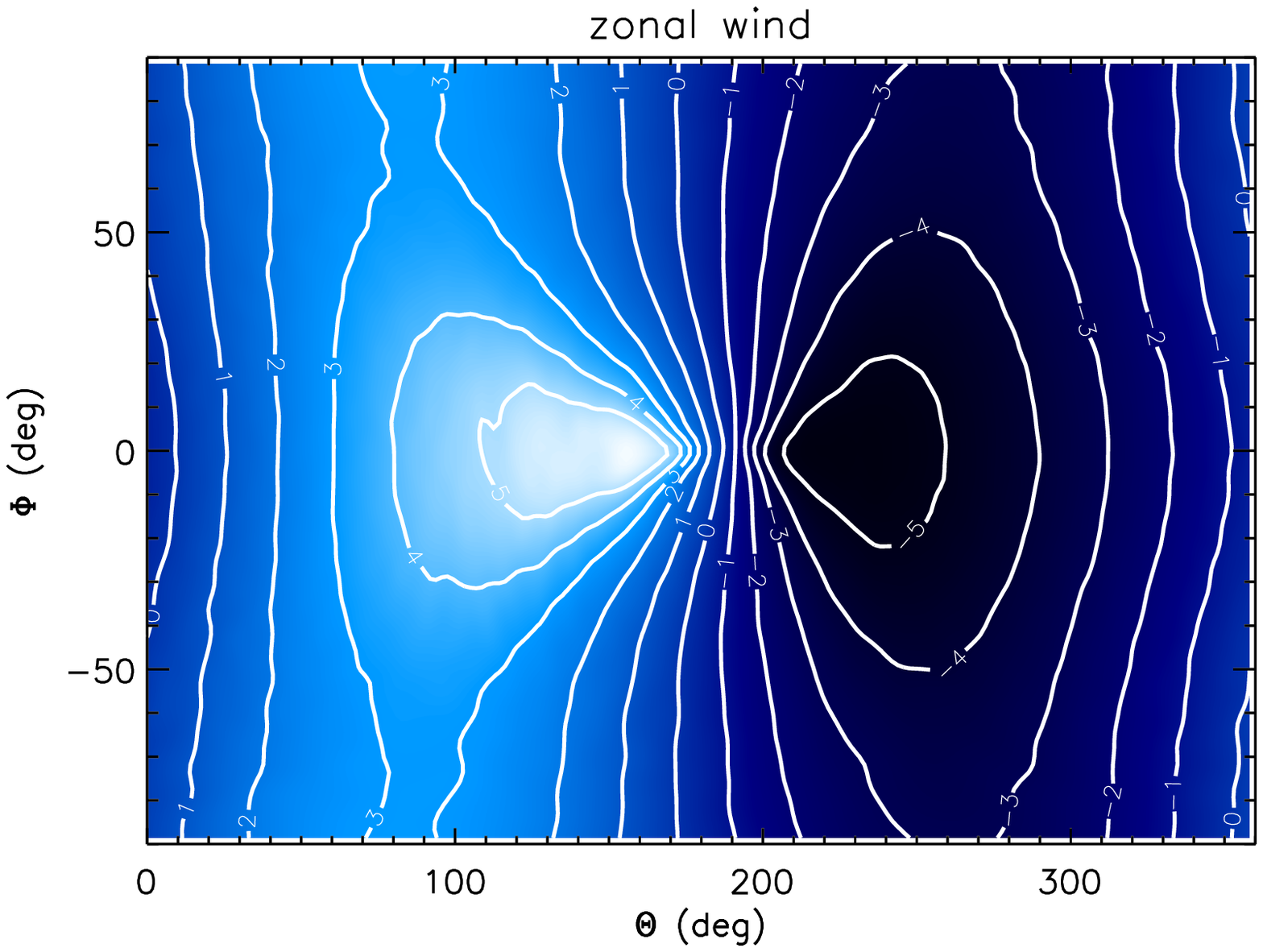}
\includegraphics[width=0.48\columnwidth]{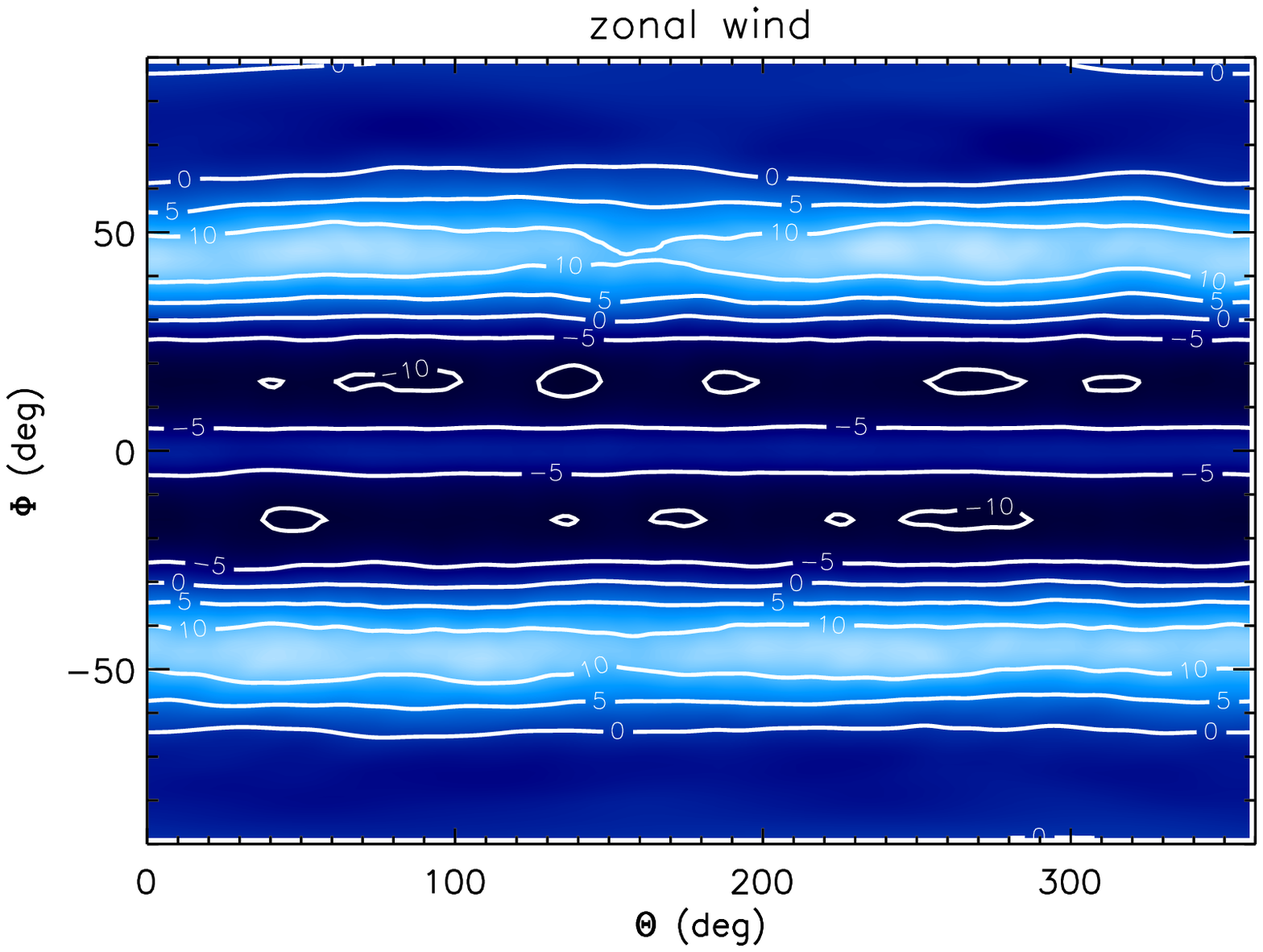}
\end{center}
\vspace{-0.2in}
\caption{Same as Figure \ref{fig:temperature} but for the long-term, global, zonal wind maps.  Contour levels are in units of m s$^{-1}$.}
\label{fig:zonal}
\vspace{-0.1in}
\end{figure}

In their simplest form, the primitive equations collectively describe a frictionless heat engine, where no viscous terms exist to convert mechanical energy back into heat.  This is operationally accomplished by the addition of a ``hyperviscous" term, which also ensures numerical stability by quenching small-scale noise accumulating at the (horizontal) grid scale (e.g., \citealt{shapiro70,rvs80,s94,mr09,hmp11}).  Following \cite{hs94}, we set the hyperviscous time scale to be 0.1 Earth days in a fourth-order hyperviscosity scheme.

Readers interested in more technical details of the model are referred to \texttt{http://www.gfdl.noaa.gov/fms}.

\subsection{Held-Suarez forcing}

The effects of stellar irradiation and geometry, known as the thermal forcing, on the atmosphere are encapsulated in the forcing function $T_{\rm force}$ \citep{hs94}.  In the classic Held-Suarez benchmark, the thermal forcing function is designed to reproduce the observed large-scale climate patterns on Earth,
\begin{equation}
T_{\rm force} = \left[ T_0 - \Delta T_{\rm EP} \sin^2\Phi - \Delta T_z \ln{\left(\frac{P}{P_0} \right)} \cos^2\Phi \right] \left( \frac{P}{P_0} \right)^\kappa,
\label{eq:ths}
\end{equation}
where $T_0=315$ K is the surface temperature at the equator (which we will scale down later), $\Delta T_{\rm EP}=60$ K is the temperature difference between the equator and the poles, $P$ represents the vertical pressure and $\Phi$ denotes the latitude.  The third term in equation (\ref{eq:ths}) is a stabilizing term where $\Delta T_z=10$ K.  Knowledge of the specific heat capacity at constant pressure $c_p$ and the ideal gas constant ${\cal R}$ allow for the specification of $\kappa \equiv {\cal R}/c_p$.  In the absence of observational constraints, we adopt terrestrial values for these quantities: $c_p = 1004.64$ J kg$^{-1}$ K$^{-1}$, ${\cal R}=287.04$ J kg$^{-1}$ K$^{-1}$ and $\kappa = 2/7$.  These values are plausible, since $c_p \sim 10^3$ J kg$^{-1}$ K$^{-1}$ for a wide variety of known gases and $\kappa=2/7$ has been adopted even for simulations of hot Jupiters (e.g., \citealt{mr09,hmp11}).  The surface pressure is initially assumed to be $P_0=1$ bar, but we will explore variations in this parameter in \S\ref{subsect:p0}.  We have initially not considered the presence of a tropopause, but this possibility is examined in \S\ref{subsect:tropo}.

Thermal forcing for a tidally-locked exoplanet can be mimicked by replacing the $-\sin^2\Phi$ term in equation (\ref{eq:ths}) with $\cos(\Theta-180^\circ) \cos\Phi$ (e.g., \citealt{cs05,cs06,mr09,ms10,hmp11}), such that the substellar point is located at $\Theta = 180^\circ$ and $\Phi=0^\circ$ with $\Theta$ denoting the longitude.  Equation (\ref{eq:ths}) can therefore be modified to become
\begin{equation}
T_{\rm force} = \left[ T^\prime_0 + \Delta T_{\rm EP} \cos\left( \Theta - 180^\circ \right) \cos \Phi - \Delta T_z \ln{\left(\frac{P}{P_0}\right)} \cos^2\Phi \right] \left( \frac{P}{P_0} \right)^\kappa,
\label{eq:tidal}
\end{equation}
where $T^\prime_0$ is now the surface temperature at the poles.  Operationally, we find that Gliese 581g simulations with $\Delta T_{\rm EP}=60$ K and assuming tidal locking do not come to quasi-equilibrium within the duration of the simulations (in the sense that the temperature map produced is discrepant from the forcing function), which is contrary to expectations because radiative cooling occurs quickly and the global temperature map should simply relax to the thermal forcing function (in the absence of initial background flow; \citealt{tc10}).  We therefore execute several simulations with different values of $\Delta T_{\rm EP}$ to investigate this issue.  We find that the \emph{global structure} of the temperature and wind maps obtained are insensitive to the choice of $\Delta T_{\rm EP}$ within the range 10--60 K, a point we will explicitly demonstrate in \S\ref{subsect:dt}.  For operational reasons, we therefore select $\Delta T_{\rm EP} = 10$ K for the tidally-locked case because the simulations do attain quasi-equilibrium.  Nevertheless, $\Delta T_{\rm EP}$ is  an unconstrained parameter of the system.

\begin{figure}
\begin{center}
\includegraphics[width=0.48\columnwidth]{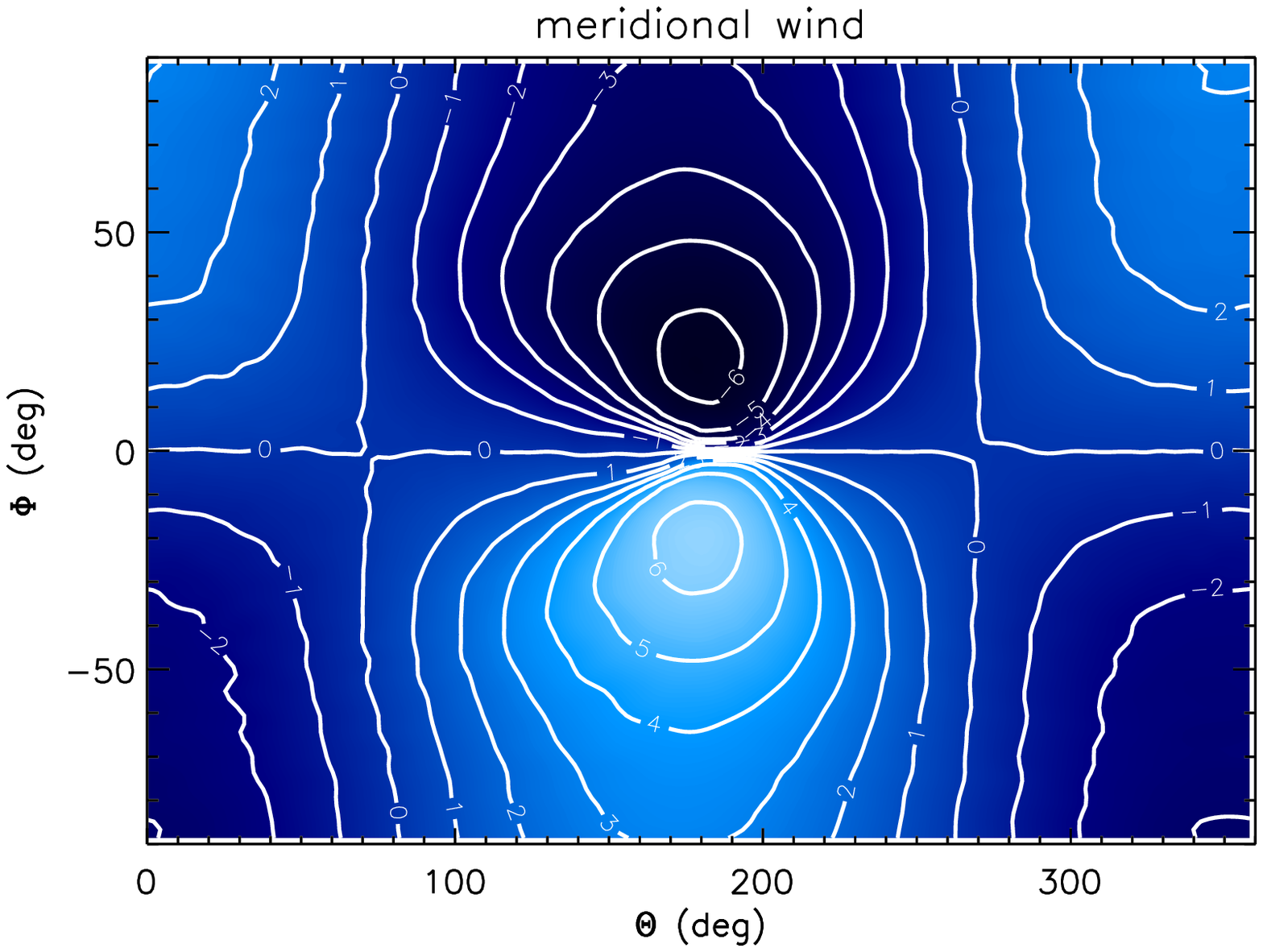}
\includegraphics[width=0.48\columnwidth]{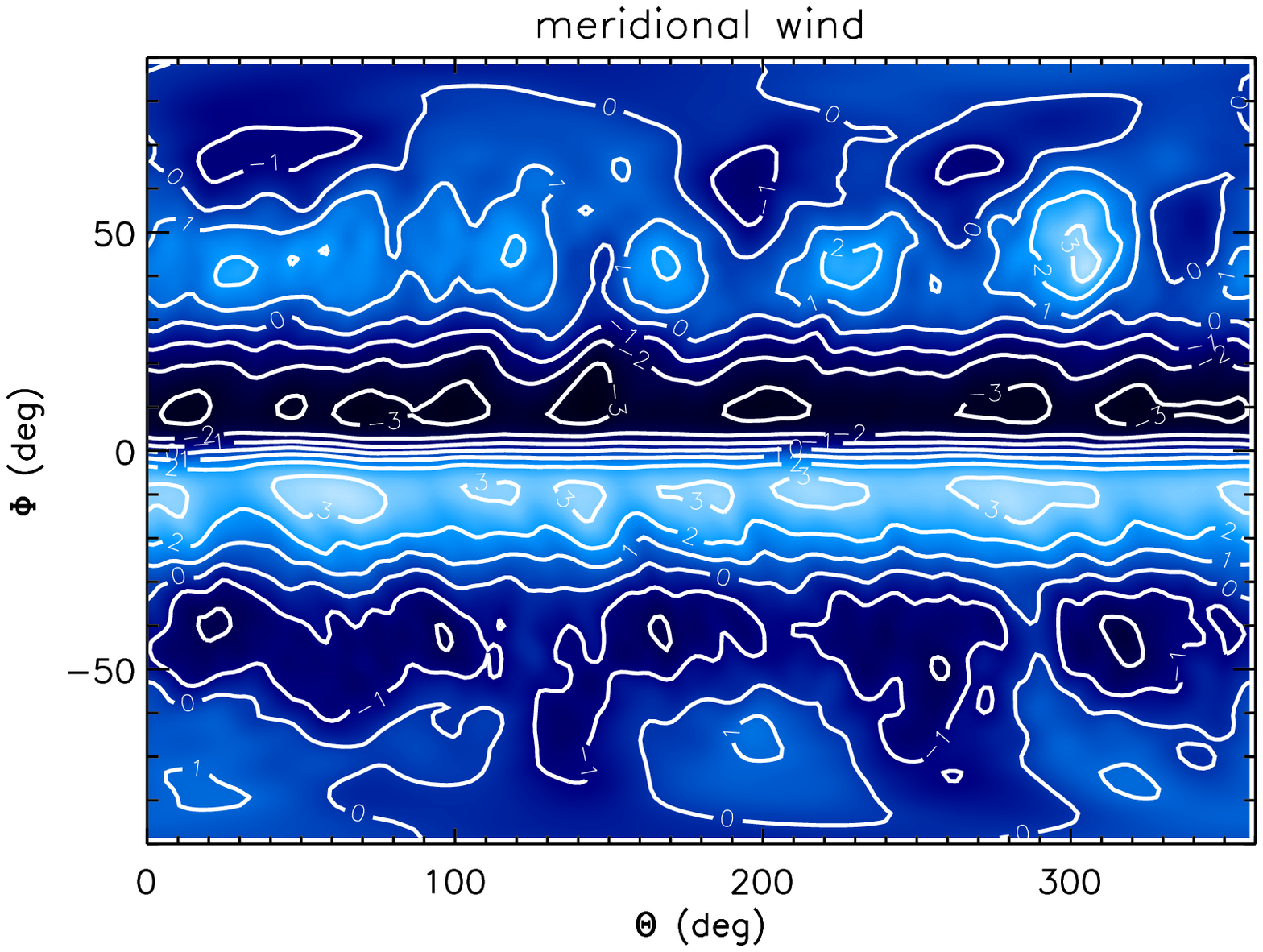}
\end{center}
\vspace{-0.2in}
\caption{Same as Figure \ref{fig:zonal} but for the long-term, global, meridional wind maps.}
\label{fig:meri}
\vspace{-0.1in}
\end{figure}

\begin{figure}
\begin{center}
\includegraphics[width=0.48\columnwidth]{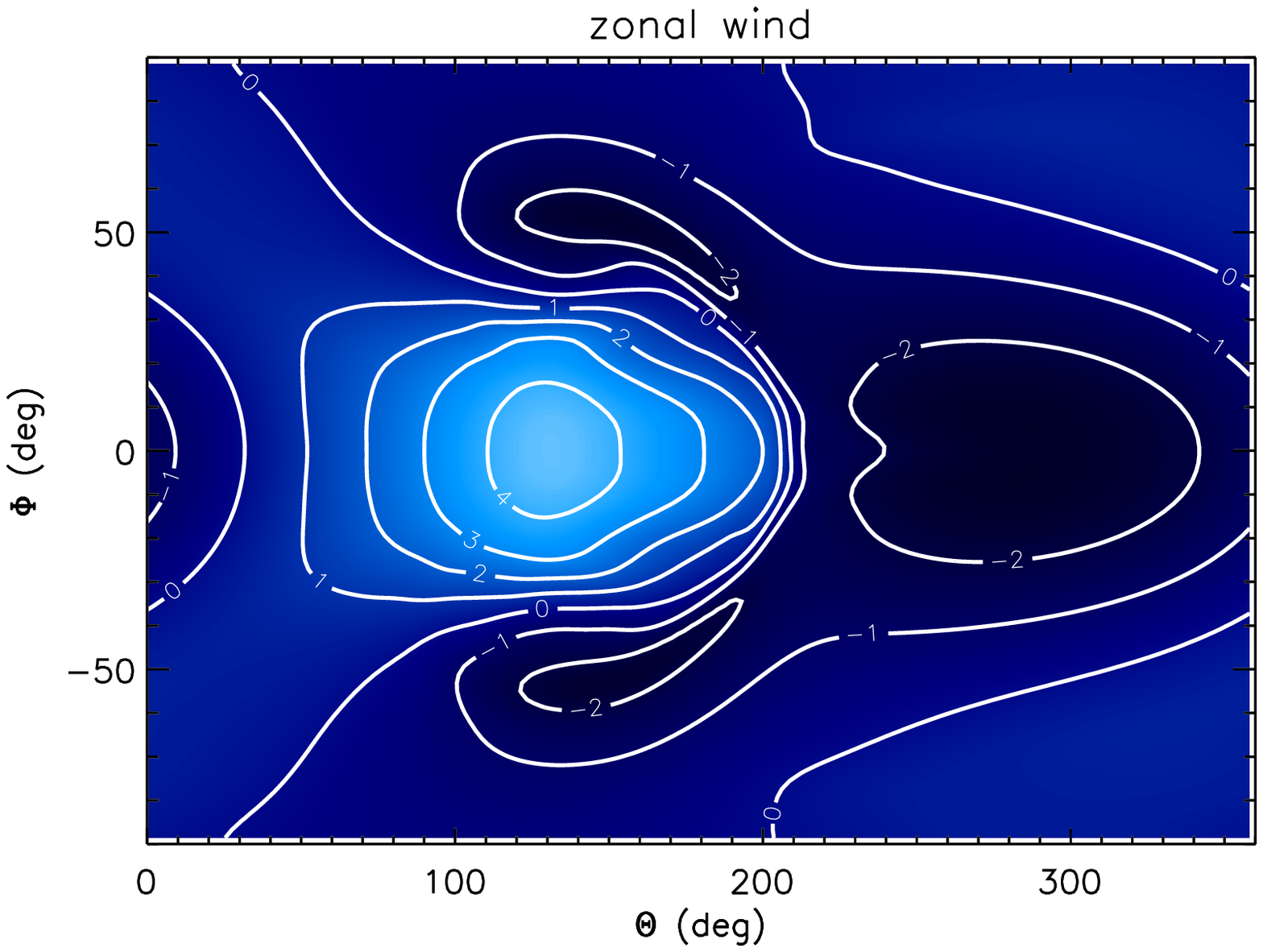}
\includegraphics[width=0.48\columnwidth]{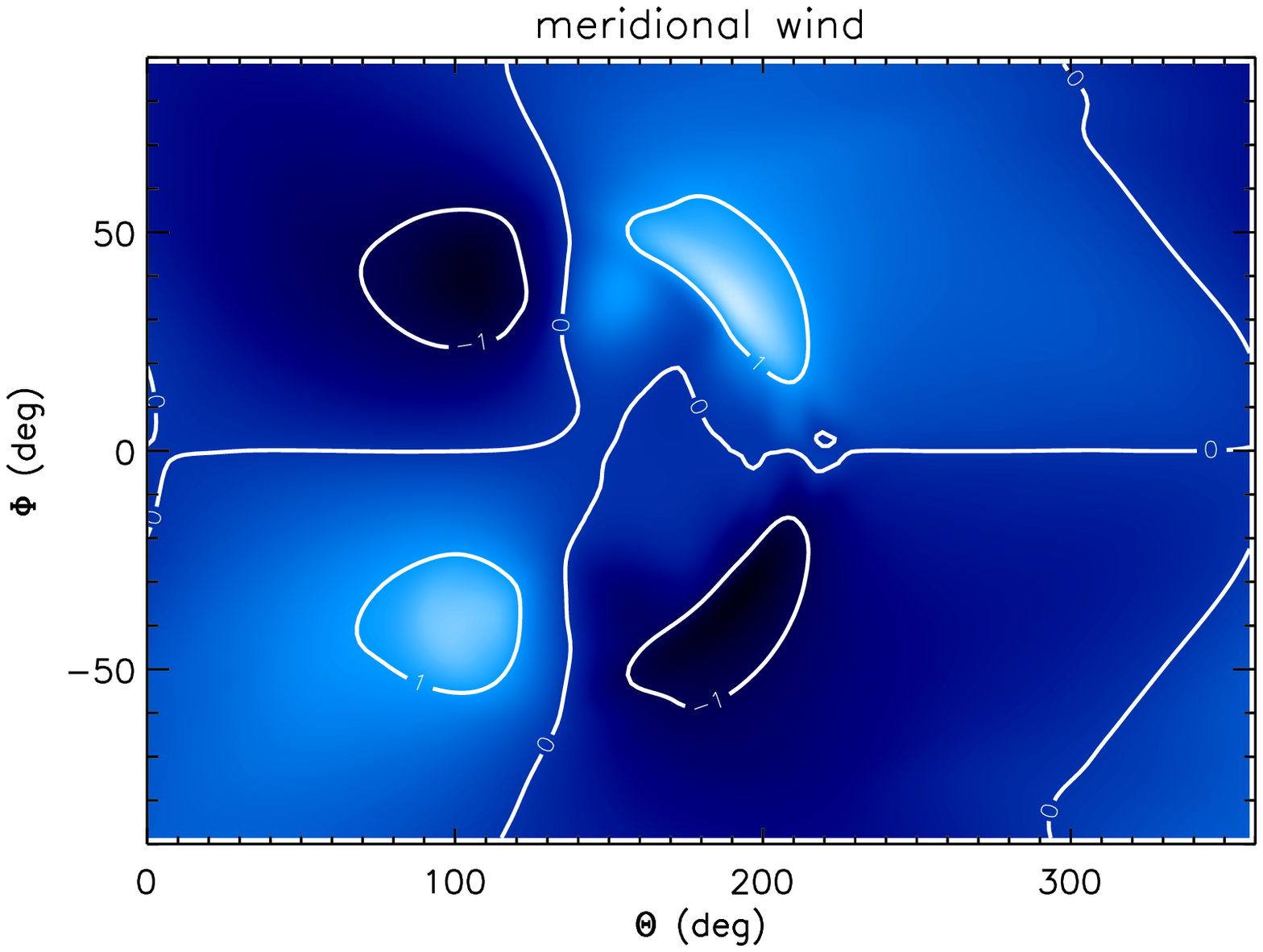}
\end{center}
\vspace{-0.2in}
\caption{Global, long-term zonal (left) and meridional (right) wind maps for a simulation assuming tidal locking, where the radiative relaxation and Rayleigh friction times are set to be 36.562 times their original values.  The maps are averaged over 1000 Earth days, where the first 2000 days of the simulation were discarded.}
\label{fig:cool}
\vspace{-0.1in}
\end{figure}

\subsection{Physical parameters: scaling}

We next scale the value of $T_0$ in equation (\ref{eq:ths}) to one appropriate to Gliese 581g.  Using the scaling,
\begin{equation}
T_0 \propto {\cal L}^{1/4} a^{-1/2},
\label{eq:scaling}
\end{equation}
where ${\cal L}$ is the stellar luminosity and $a$ is the distance from the star, it follows that $T_0 = 278$ K since Gliese 581 has a luminosity of $0.013~{\cal L}_\odot$ and $a=0.14601$ AU \citep{mayor09,vogt10}.  Using this scaled value of $T_0$ and $\Delta T_{\rm EP}=10$ K, we get $T^\prime_0 = 268$ K.  It is important to note that $T_0$ is \emph{not} the ``equilibrium temperature" (i.e., blackbody equivalent) of the exoplanet \citep{sel07}, which is estimated to be $T_{\rm eq} \approx 230$ K (assuming a Bond albedo of 0.3, typical for Solar System objects; \citealt{vogt10}).  It is also important to note that typical estimates of $T_{\rm eq}$ assume that energy is not transported from the permanent day to the night side (assuming tidal locking) of the exoplanet.  On Earth, the equilibrium temperature is about 280 K, which is lower than the surface temperature of 315 K assumed in the Held-Suarez benchmark.  The increased surface temperature is due to a combination of a non-zero surface albedo and the retention of (re-emitted) longwave radiation by greenhouse gases (mostly carbon dioxide and water vapour).  Furthermore, the temperature difference between the equator and the poles is determined by a combination of atmospheric dynamics and seasonal variations (e.g., \citealt{vallis06}), so the simple scaling in equation (\ref{eq:scaling}) cannot be straightforwardly applied to $\Delta T_{\rm EP}$.  Therefore, we retain $\Delta T_{\rm EP}=60$ K in the case of a hypothetical Gliese 581g with a rotational period of one Earth day.

Only the minimum mass of Gliese 581g is currently known, but the dynamical stability analyses of \cite{mayor09} and \cite{vogt10} --- assuming co-planar orbits --- restrict $1/\sin{i}$ to have upper limits of 1.6 and 1.4, respectively.  The mass of the exoplanet affects the assumed values of the exoplanetary radius,
\begin{equation}
R_p = \left( \frac{M_p}{M_\oplus} \right)^{1/3} \left( \frac{\rho_p}{ \rho_\oplus} \right)^{-1/3} R_\oplus,
\label{eq:radius}
\end{equation}
and the surface gravity,
\begin{equation}
g_p = \left( \frac{M_p}{M_\oplus} \right)^{1/3} \left( \frac{\rho_p}{\rho_\oplus} \right)^{2/3} g_\oplus,
\label{eq:grav}
\end{equation}
where $R_\oplus = 6371$ km and $g_\oplus = 980$ cm s$^{-2}$.  If $M_p = 3.1~M_\oplus$, then the scaling factor associated only with the mass is 1.46.  If we instead have $M_p = 4.3~M_\oplus$ and $5.0~M_\oplus$, then the scaling factors are 1.63 and 1.71, respectively.

In the spirit of a scaled-up Earth, we initially assume Gliese 581g to have the same mass density as Earth ($\rho_p/\rho_\oplus=1$) and take the minimum mass to be the actual mass, $M_p = 3.1~M_\oplus$, but explore the implications of raising the assumed mass to $M_p=4.3~M_\oplus$ and $5.0~M_\oplus$ in \S\ref{subsect:mass}.  In \S\ref{subsect:rho}, we vary the assumed value of $\rho_p/\rho_\oplus$ and explore the implications.  We may already anticipate that the flow structure is mainly determined by the rotational rate ($\Omega_p$) and the details of the gas physics (the frictional and radiative cooling times).

The observed orbital period of Gliese 581g is 36.562 Earth days \citep{vogt10}.  If the exoplanet is tidally-locked, then the angular rotational frequency is $\Omega_p = 1.989 \times 10^{-6}$ s$^{-1}$.  In \S\ref{subsect:tcool}, we explore the implications of varying the global radiative cooling time for a tidally-locked Gliese 581g --- in essence, we are quantifying the relative importance of cooling versus advection.  If we relax the assumption of tidal locking, then we can explore the effect of varying the rotational frequency of the exoplanet (\S\ref{subsect:rotate}).  For example, if Gliese 581g has a rotational period of one Earth day, then $\Omega_p = 7.292 \times 10^{-5}$ s$^{-1}$.

\section{Results}
\label{sect:results}

We begin by presenting results from a trio of baseline models, which we consider to be representative.  Subsequently, we present a suite of simulations in which we systematically vary a set of physical parameters in order to determine which of them are significant.  We conclude that the major parameters are the radiative cooling time $\tau_{\rm rad}$ and the rotational frequency $\Omega_p$.

\subsection{Baseline models}
\label{subsect:baseline}

\begin{figure}
\begin{center}
\includegraphics[width=0.6\columnwidth]{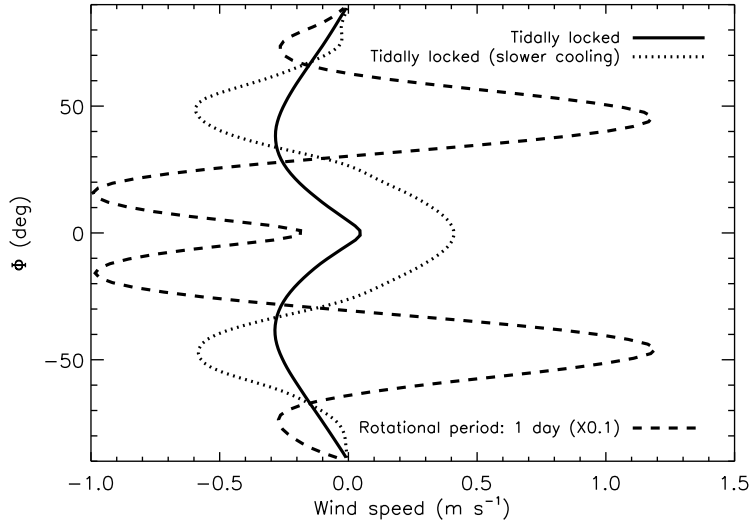}
\end{center}
\vspace{-0.2in}
\caption{Zonally- and temporally-averaged zonal wind speeds near the surface of Gliese 581g ($P=0.95$--1 bar).  Positive and negative values indicate super- and counter-rotation, respectively.}
\label{fig:wind}
\vspace{-0.1in}
\end{figure}

The left panel of Figure \ref{fig:mollweide} shows the Mollweide projection\footnote{Pseudo-cylindrical projection of a globe which conserves area but not angle or shape.  Also called the ``homalographic projection".} of a snapshot from the simulation where Gliese 581g is assumed to have a mass of $M_p=3.1~M_\oplus$, a mass density of $\rho_p = \rho_\oplus$, a surface pressure of $P_0 = 1$ bar and is tidally-locked.  The Newtonian cooling time is about 4 days, following the Held-Suarez benchmark for Earth (see \S\ref{subsect:tcool}).  Since the rotational period of about 37 days is much longer than the radiative cooling time, the structure of the flow is sculpted by radiation rather than advection.  The flow may also be modified by unresolved gravity waves \citep{wc10}.  The relatively fast cooling time implies that the global temperature map relaxes approximately to the input thermal forcing function.

While such visualizations are aesthetically pleasing, more insight is provided by looking at the temporally-averaged temperature and wind maps as functions of longitude and latitude --- the long-term, quasi-stable climate.  This is shown in Figure \ref{fig:temperature}, where we contrast both the tidally-locked and non-tidally-locked cases.  For the tidally-locked case, the permanent day side of the exoplanet is just within the classical $T=0^\circ$--$100^\circ$C habitable temperature range.  In the non-tidally-locked case where the rotational period is assumed to be equal to one Earth day, the flow is dominated by advection rather than radiation with the temperatures at the equator hovering around a few degrees Celsius.  The pair of global temperature maps in Figure \ref{fig:temperature} makes the point that conclusions on the exact locations for habitability on the surface of an exo-Earth depend upon whether the assumption of tidal locking is made (see also \S\ref{subsect:rotate}).  Even on the cold night side, the temperatures are comparable to those experienced in Antarctica where colonies of algae have been discovered and analyzed \citep{ed04,ed04b}.  All of these statements are made keeping in mind that temperature is a necessary but insufficient condition for habitability (see \S\ref{sect:discussion}).

Figures \ref{fig:zonal} and \ref{fig:meri} show the global zonal and meridional wind maps, respectively.  In the case of a tidally-locked Gliese 581g, fluid is transported across hemispheric scales at speeds $\sim 1$ m s$^{-1}$, comparable to typical wind speeds on Earth.  The wind patterns have a slight asymmetry from west to east due to the rotation of the exoplanet.  If the exoplanet instead has a rotational period of one Earth day (and is not tidally locked), there is longitudinal homogenization of the winds with a counter-rotating jet at the equator and super-rotating jets at mid-latitude.  The meridional wind map is now characterized by smaller structures.  The slightly faster wind speeds recovered from the simulation with a rotational period of one Earth day are artifacts of assuming a higher value of $\Delta T_{\rm EP}$ (60 K versus 10 K)--- nevertheless, the \emph{global structure} of the wind maps are robust predictions of the simulations, which we will demonstrate in \S\ref{subsect:dt}.

To further explore the interplay between radiative cooling and advection, we execute another simulation where the radiative cooling (originally 4 Earth days) and Rayleigh friction (originally 1 Earth day) times are set to be 36.562 times their fiducial values --- in essence, we are scaling by the ratio of the observed orbital period of Gliese 581g to the rotational period of Earth.\footnote{As an aside, we note that scaling down from Venus is an alternative, plausible approach since the Venusian orbital period is about 117 days.}  Basically, we are exploring the possibility that there is an unidentified cooling mechanism with a time scale that scales as the rotational period (assuming tidal locking).  The key point is not in the exact value of this lengthening of the cooling time, but that it may well exceed the rotational period on a tidally-locked exoplanet.  Due to the longer cooling time assumed, we now run the simulation for 3000 Earth days and discard the first 2000 days so as to attain quasi-equilibrium.  The Mollweide snapshot of the temperature and velocity fields, as well as the long-term wind maps, are shown in the right panel of Figure \ref{fig:mollweide} and also Figure \ref{fig:cool}.  Since advection occurs somewhat faster than radiative cooling, zonal winds on the exoplanetary surface develop a stronger east-west asymmetry and there are hints of energy transport from the permanent day to the night side.  The chevron-shaped feature residing around the substellar point is reminiscent of that seen at $\sim 0.1$ bar in 3D atmospheric circulation simulations of hot Jupiters (e.g., \citealt{mr09,hmp11}).  Trailing the feature are large-scale vortices spanning about a third of the hemisphere in size --- their large sizes are a consequence of the Rossby deformation length scale being relatively larger due to the slower rotation of the exoplanet when tidally locked \citep{cho03}.  In Figure \ref{fig:wind}, we show the zonally- and temporally-averaged zonal wind speeds obtained from our trio of baseline models.  The models with tidal locking both have equatorial, super-rotating winds, where a longer radiative cooling time leads to faster speeds because of the increased effectiveness of advection.  The non-tidally-locked model which assumes a rotational period of one Earth day has a counter-rotating wind at the equator; its faster speed is again an artifact of assuming a larger value of $\Delta T_{\rm EP}$.  Similar to how the Rossby length scale is increased because of slower rotation, the larger Rhines scale \citep{showman10} in the tidally-locked models results in broader jets across latitude.

The trio of simulations presented here capture the essential physics of our simple model.  In the sub-sections that follow, we will explore multiple variations on a theme, so as to ascertain the major and minor physical parameters involved in determining the atmospheric dynamics of an exo-Earth.  In doing so, we will explicitly demonstrate some of the statements made in this sub-section.

\subsection{Varying the exoplanetary mass $M_p$}
\label{subsect:mass}

\begin{figure}
\begin{center}
\includegraphics[width=0.48\columnwidth]{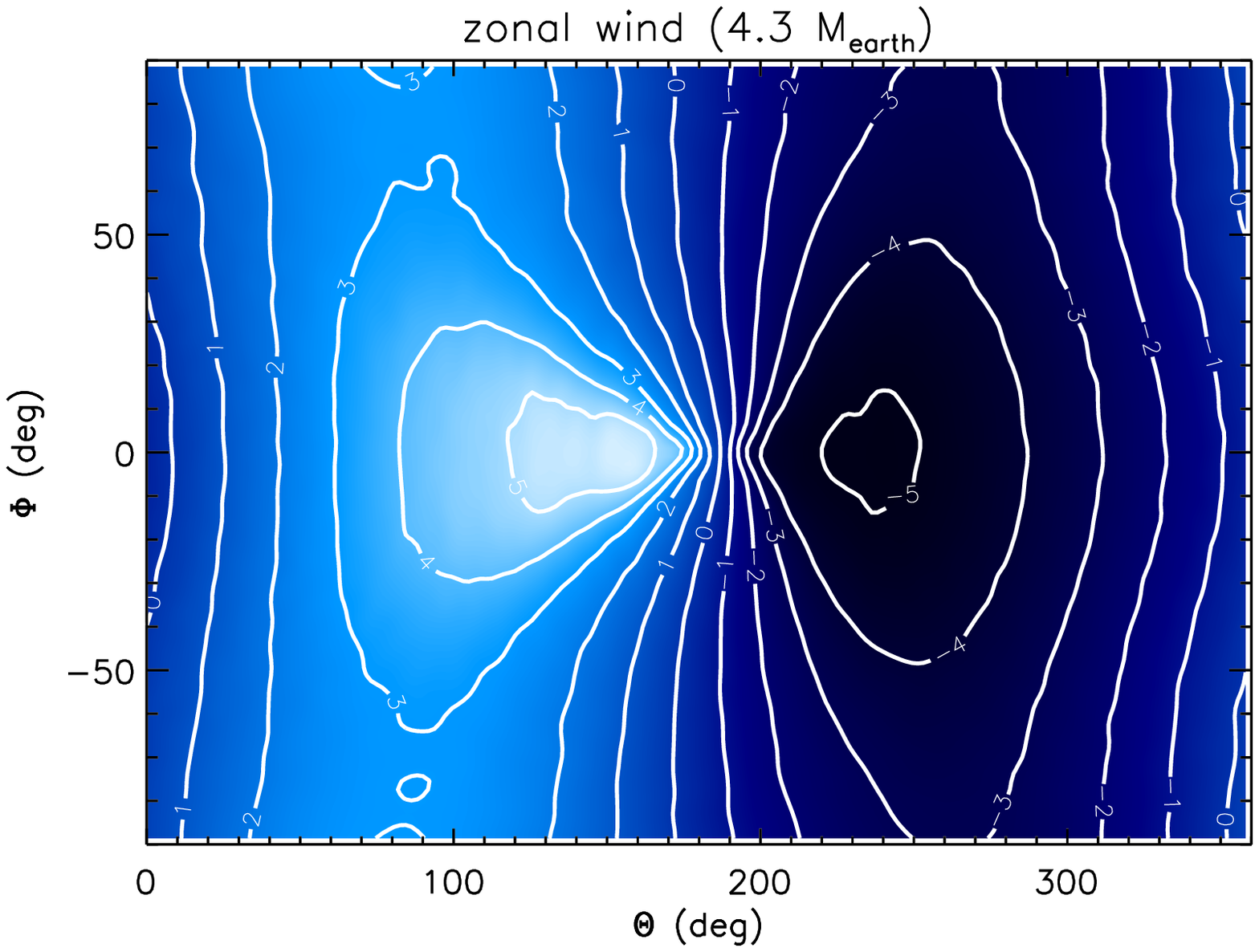}
\includegraphics[width=0.48\columnwidth]{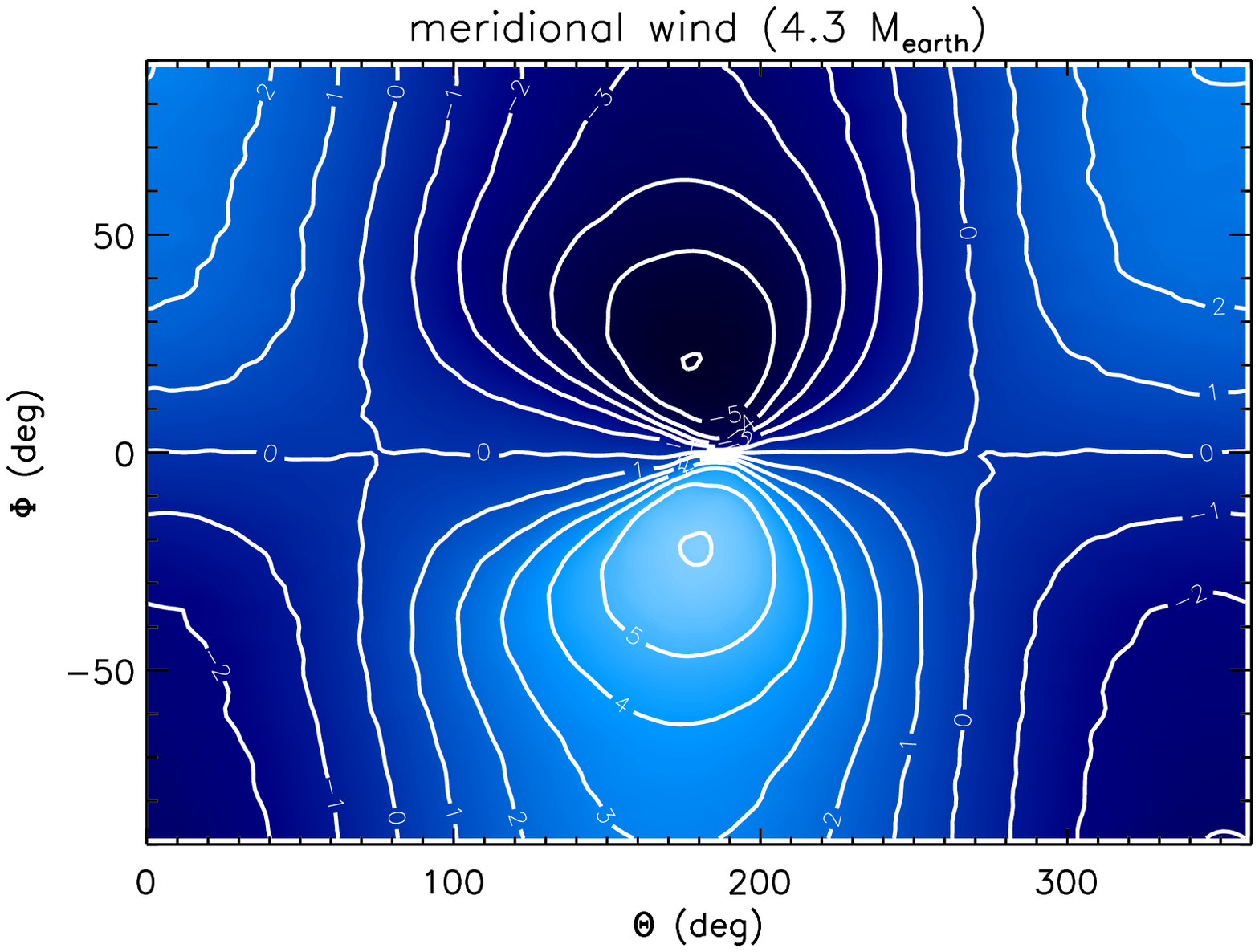}
\includegraphics[width=0.48\columnwidth]{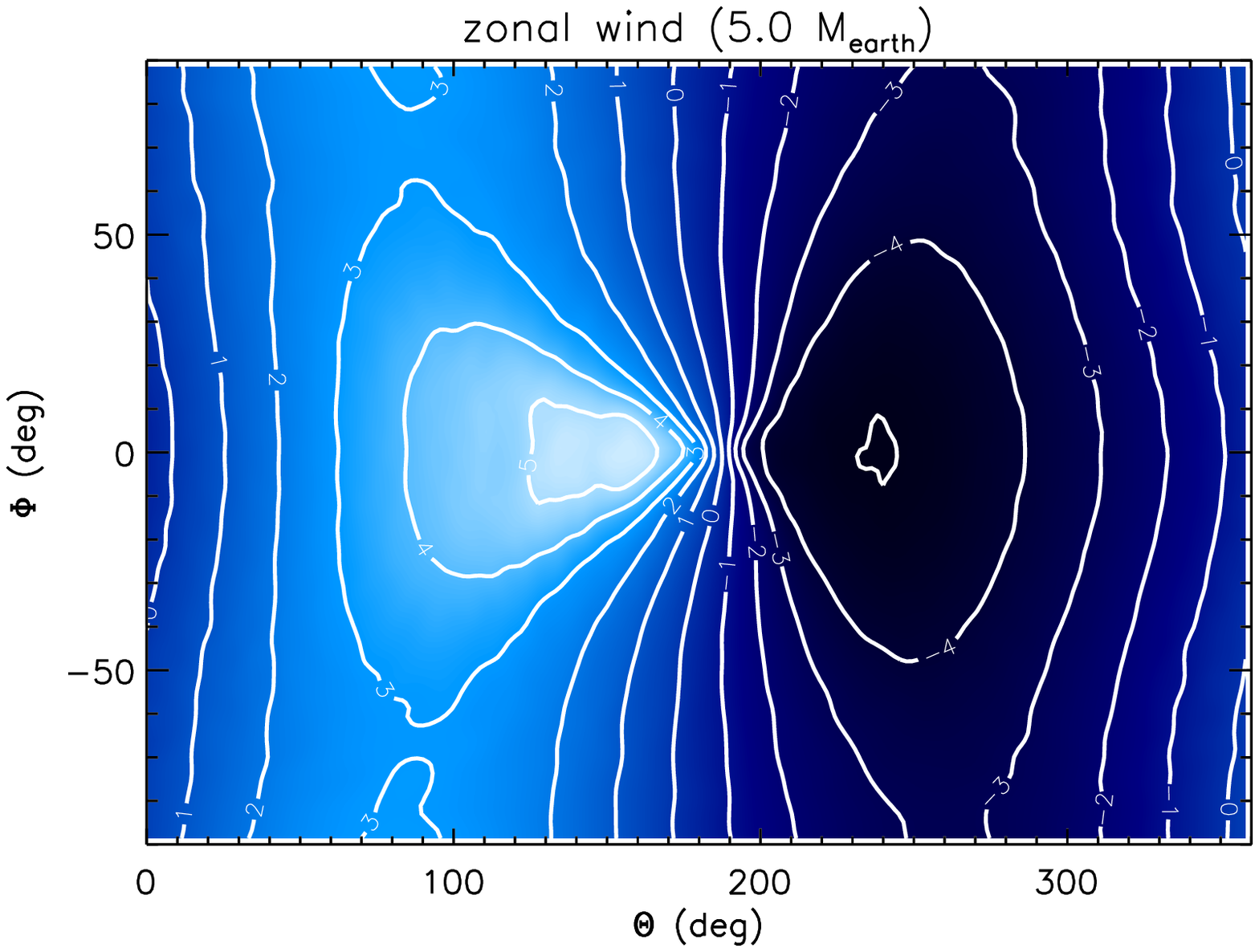}
\includegraphics[width=0.48\columnwidth]{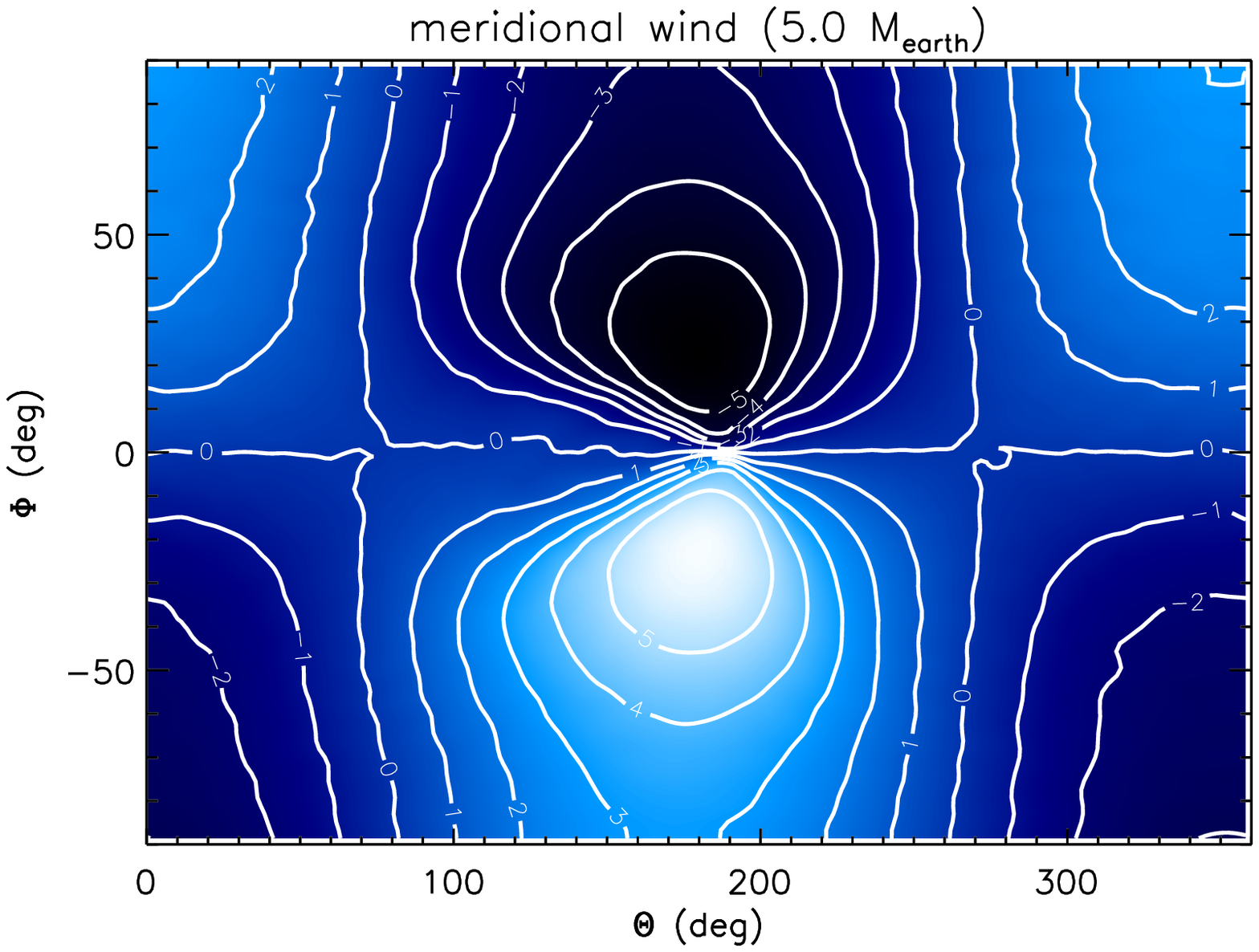}
\end{center}
\vspace{-0.2in}
\caption{Same as Figures \ref{fig:zonal} and \ref{fig:meri} but for the tidally-locked case with $M_p = 4.3~M_\oplus$ (top two panels) and $M_p = 5.0~M_\oplus$ (bottom two panels).}
\label{fig:mass}
\vspace{-0.1in}
\end{figure}

\begin{figure}
\begin{center}
\includegraphics[width=0.48\columnwidth]{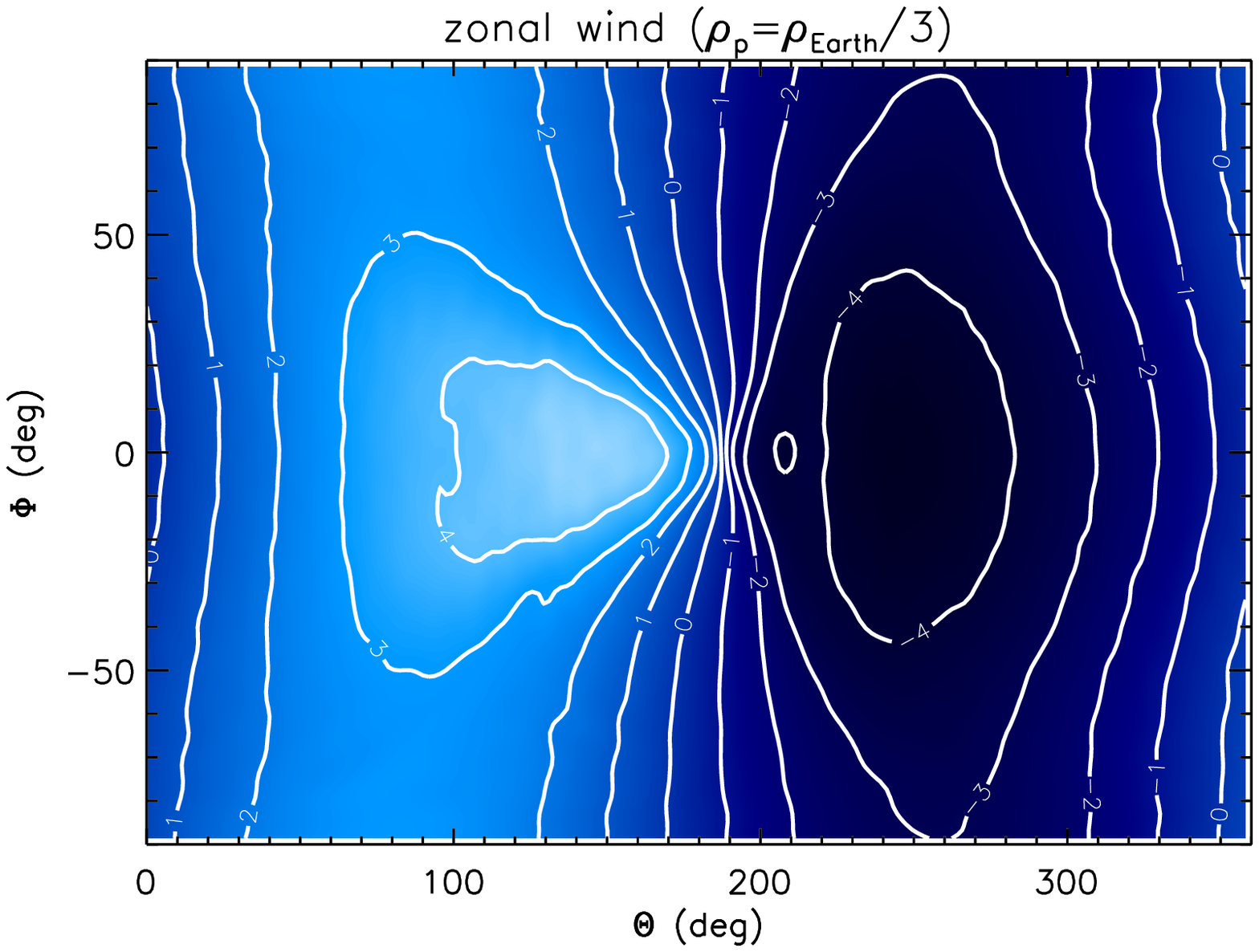}
\includegraphics[width=0.48\columnwidth]{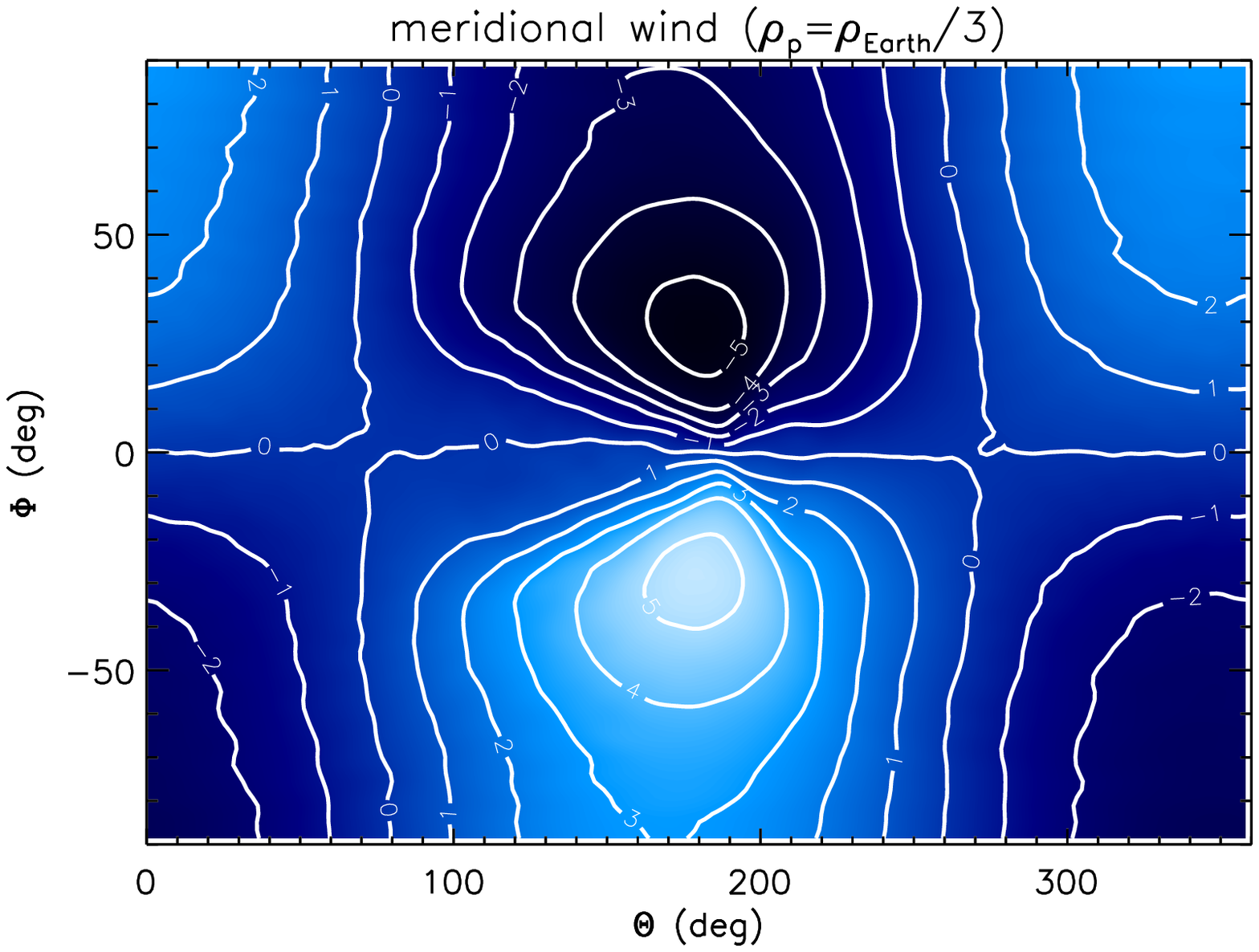}
\includegraphics[width=0.48\columnwidth]{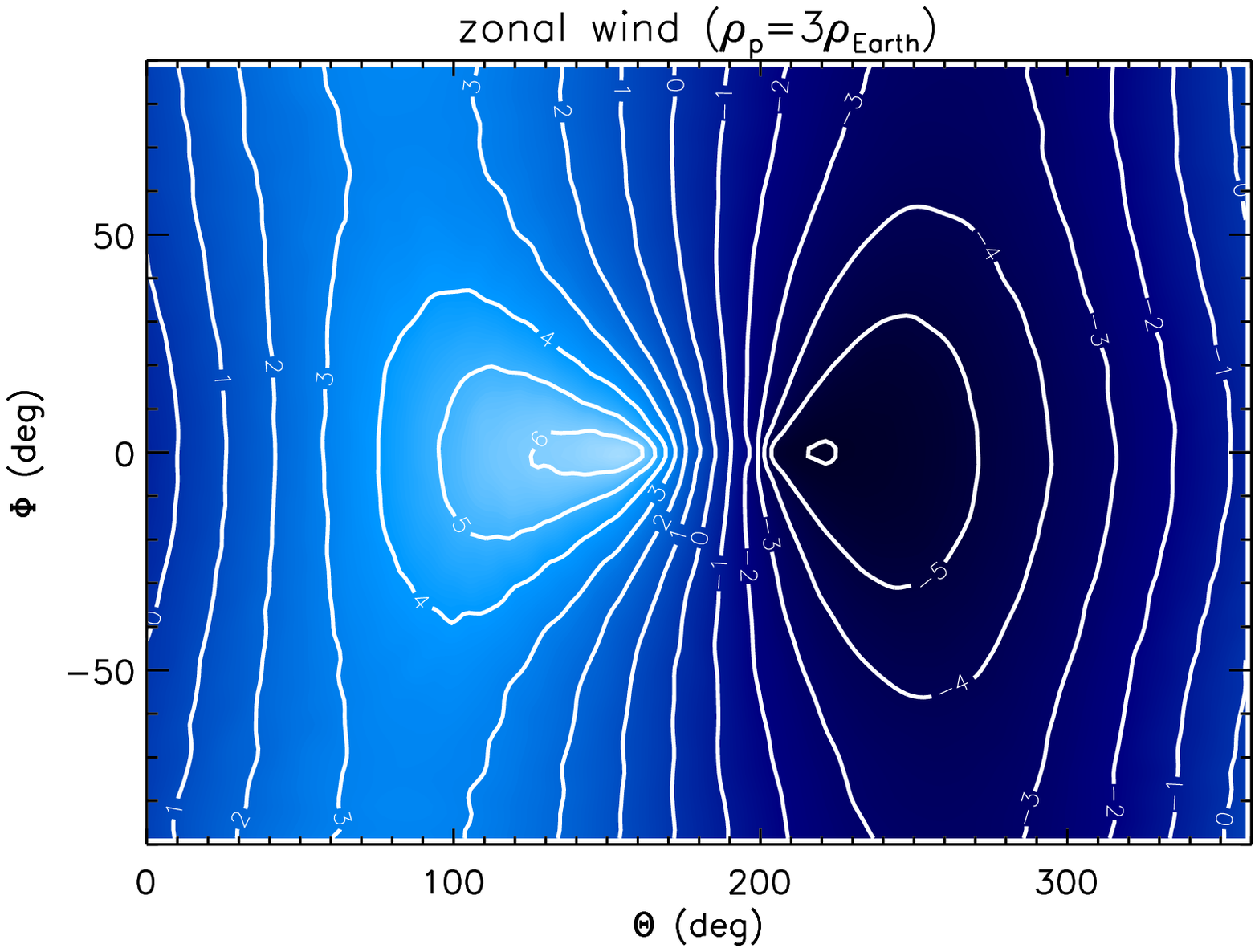}
\includegraphics[width=0.48\columnwidth]{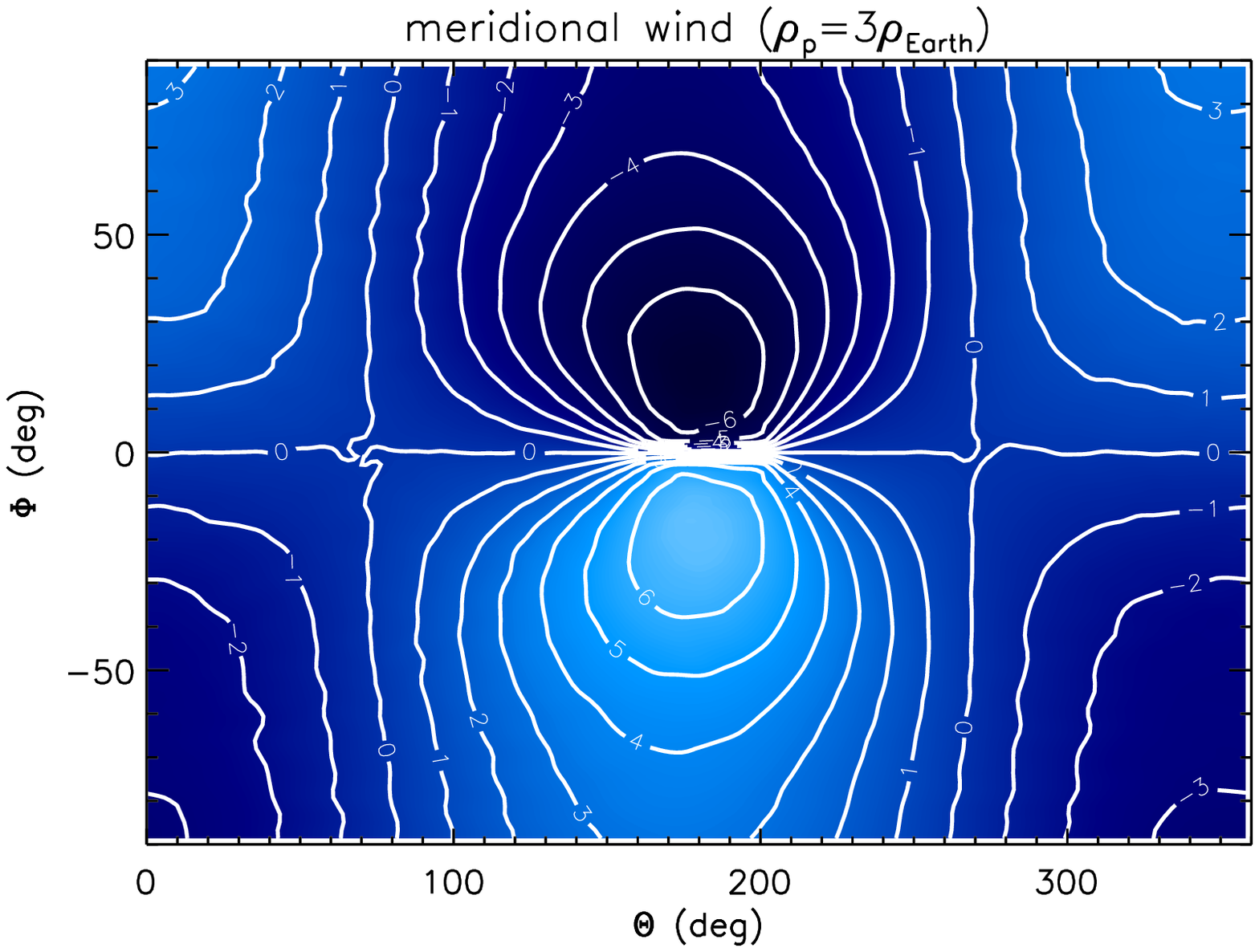}
\end{center}
\vspace{-0.2in}
\caption{Same as Figure \ref{fig:mass}, but for the tidally-locked case with $\rho_p = \rho_\oplus/3$ (top two panels) and $\rho_p = 3\rho_\oplus$ (bottom two panels).}
\label{fig:rho}
\vspace{-0.1in}
\end{figure}

Assuming the exoplanets orbiting Gliese 581 to be in coplanar orbits, \cite{mayor09} and \cite{vogt10} were able to set upper limits on $M_p$ of $4.3~M_\oplus$ and $5.0~M_\oplus$, respectively, using stability analyses that considered 4 and 6 orbiting exoplanets.  We thus explore the implications of varying $M_p/M_\oplus$ while keeping $\rho_p/\rho_\oplus = 1$.  As is evident from equations (\ref{eq:radius}) and (\ref{eq:grav}), raising the exoplanetary mass to the (maximum) values implied by the stability analyses increases both the radius and the surface gravity, but in these cases by less than a factor of 2.  As shown in Figure \ref{fig:mass}, it is therefore unsurprising that while varying $M_p$ yields minor quantitative differences in the global wind maps near the surface ($P=0.95$--1 bar), the qualitative features are largely invariant to $M_p$.

\subsection{Varying the exoplanetary mass density $\rho_p$}
\label{subsect:rho}

Varying the exoplanetary mass density is slightly different from varying the mass.  Higher exoplanetary masses generally result in larger radii and higher surface gravities.  However, a denser exoplanet has a smaller radius but higher surface gravity (and vice versa).  To gain some intuition for the possible range of mass densities involved, we note that Kuiper Belt objects have typical densities $\sim 2$ g cm$^{-3}$ \citep{fb10}, while the gas giants in our Solar System have densities $\sim 1$ g cm$^{-3}$.  By contrast, the Earth has a density of about 5.5 g cm$^{-3}$.  It is unlikely that $\rho_p$ can vary by an order of magnitude or more.  Thus, we explore the implications when $\rho_p/\rho_\oplus = 1/3$ and 3 (i.e., half an order of magnitude lower or higher than $\rho_\oplus$) and $M_p = 3.1~M_\oplus$.

Even with marked differences in the radii (6441 km versus 13397.9 km) and surface gravities (690 cm s$^{-2}$ versus 2970 cm s$^{-2}$) between the pair of simulations,  the global wind maps in Figure \ref{fig:rho} demonstrate that the mass density $\rho_p$ of the exoplanet is a minor parameter of the system.

\subsection{Varying the equator-pole temperature difference $\Delta T_{\rm EP}$}
\label{subsect:dt}

\begin{figure}
\begin{center}
\includegraphics[width=0.48\columnwidth]{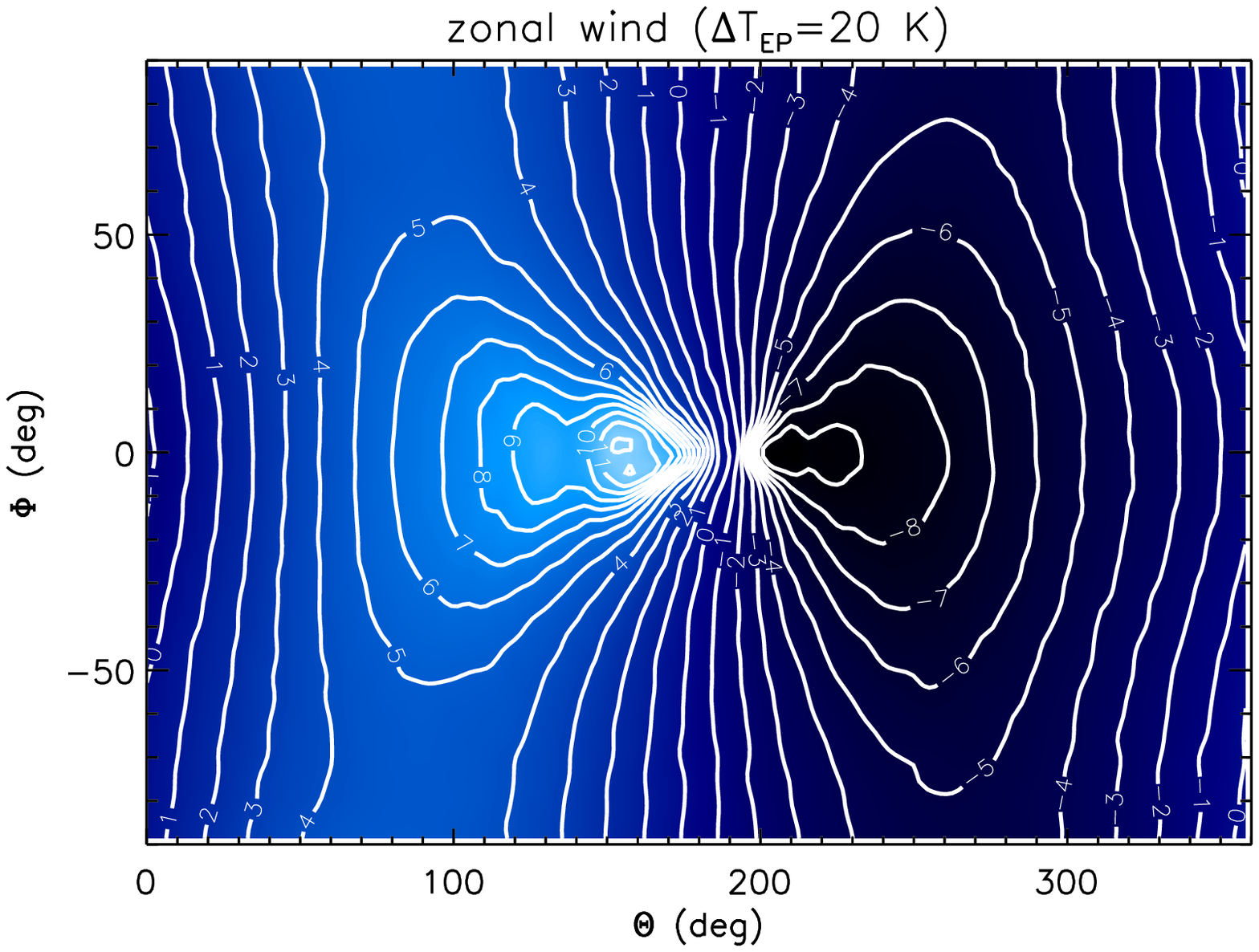}
\includegraphics[width=0.48\columnwidth]{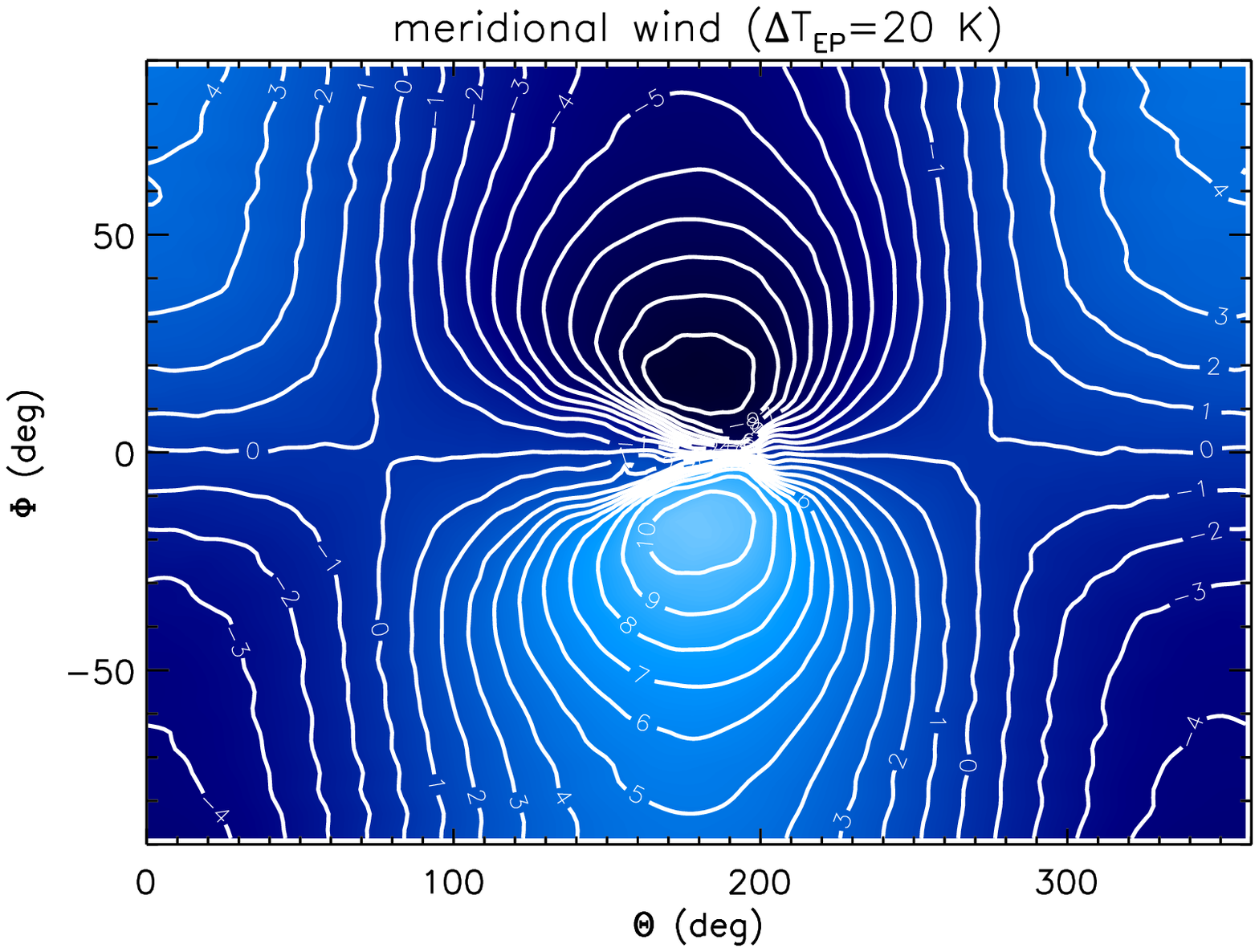}
\includegraphics[width=0.48\columnwidth]{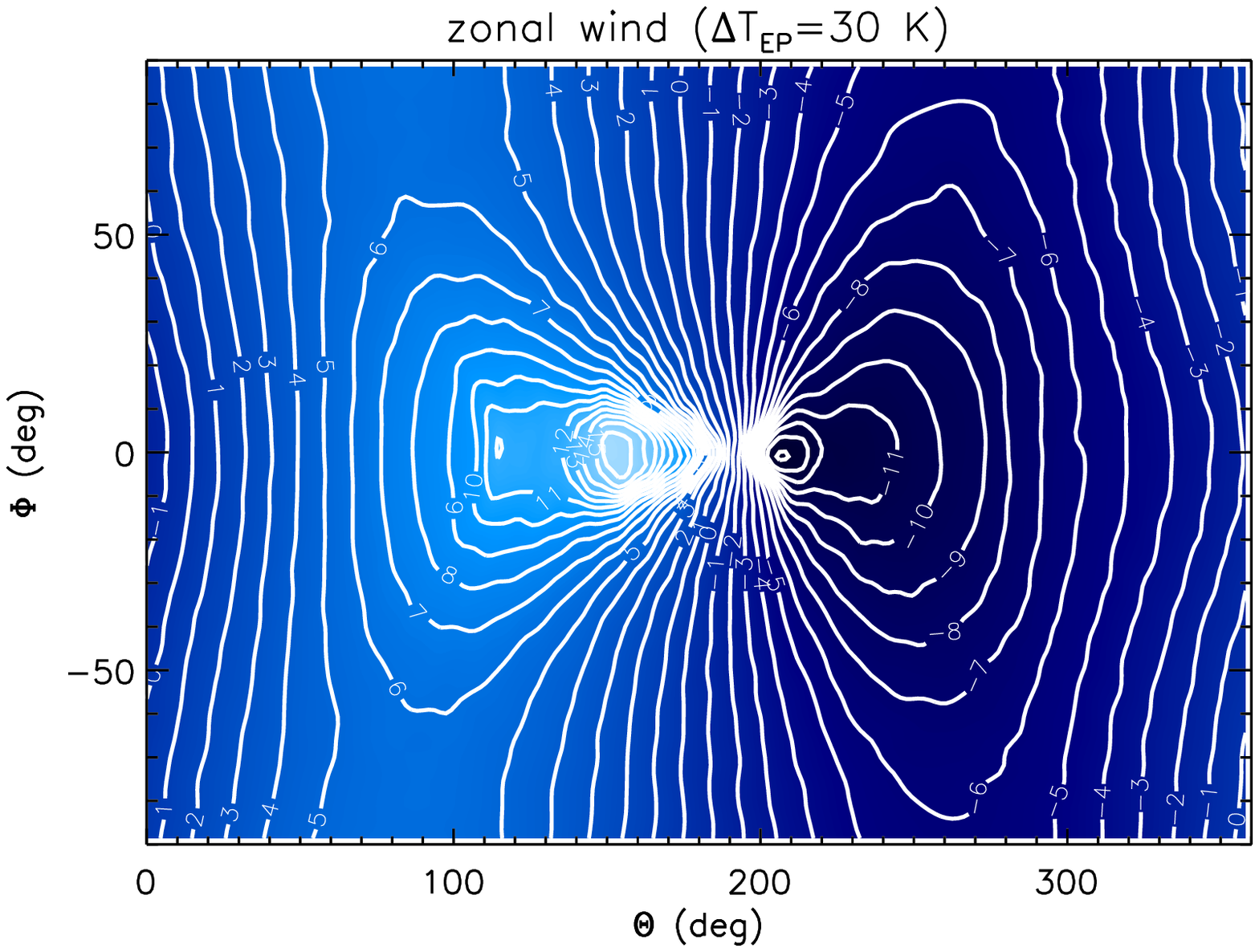}
\includegraphics[width=0.48\columnwidth]{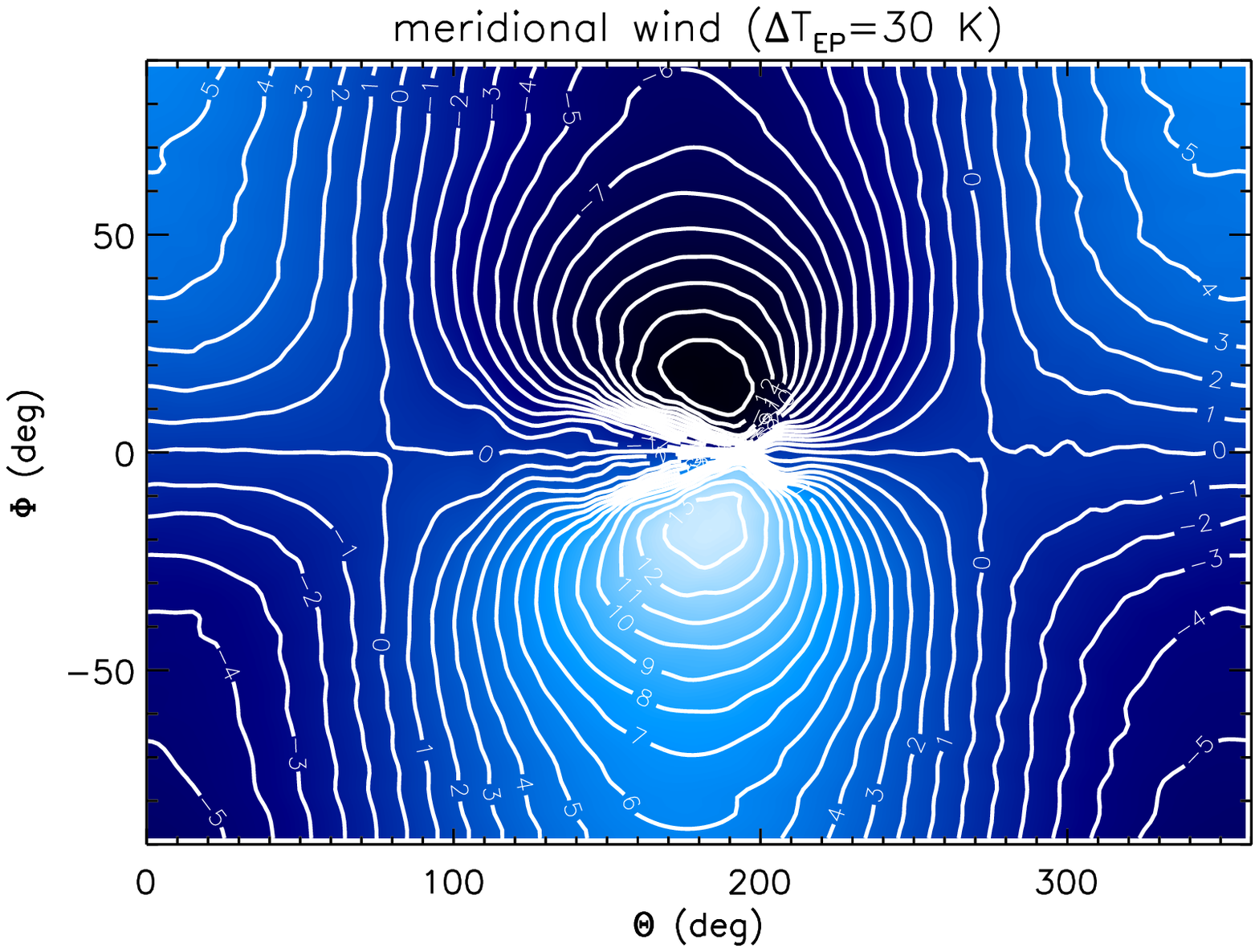}
\includegraphics[width=0.48\columnwidth]{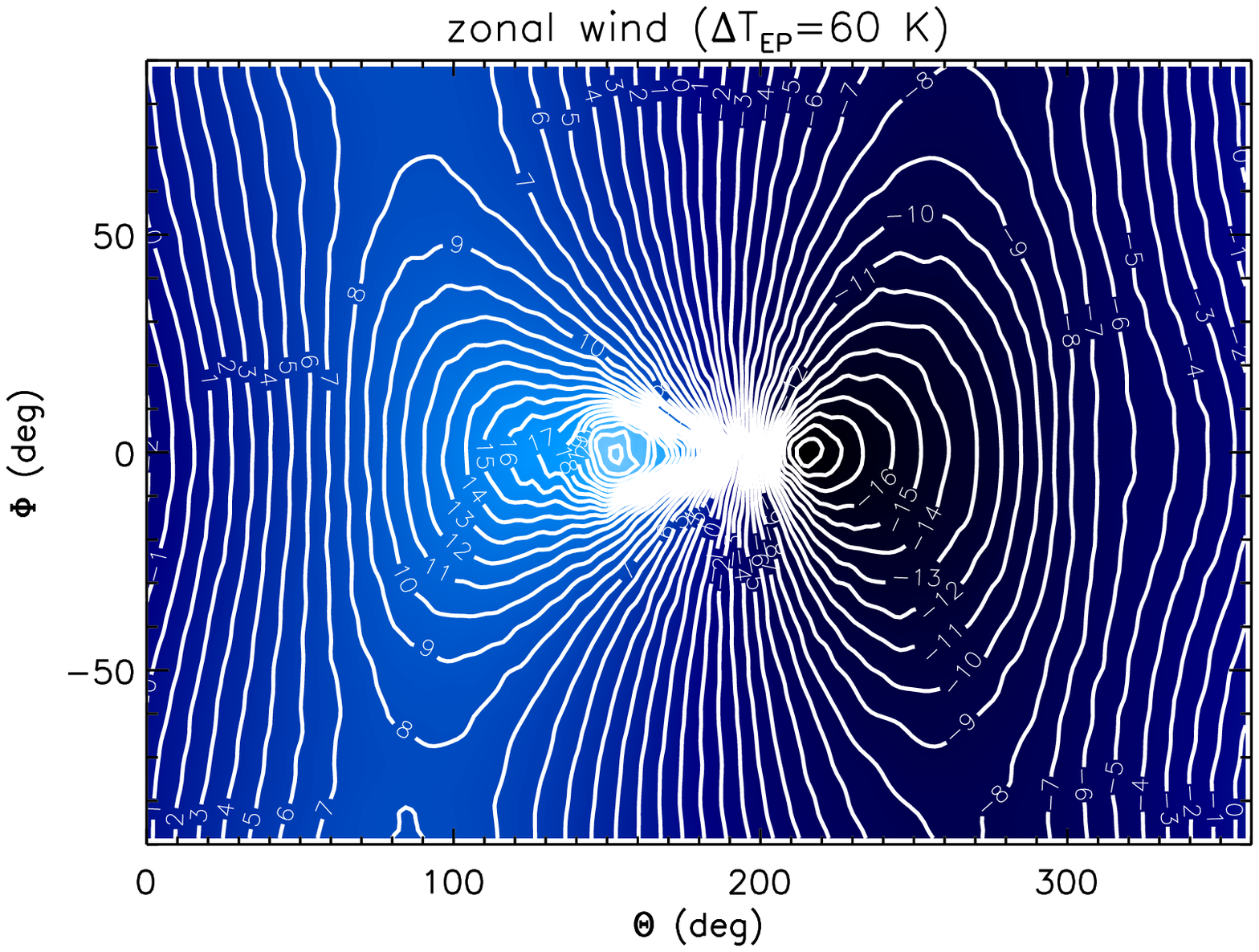}
\includegraphics[width=0.48\columnwidth]{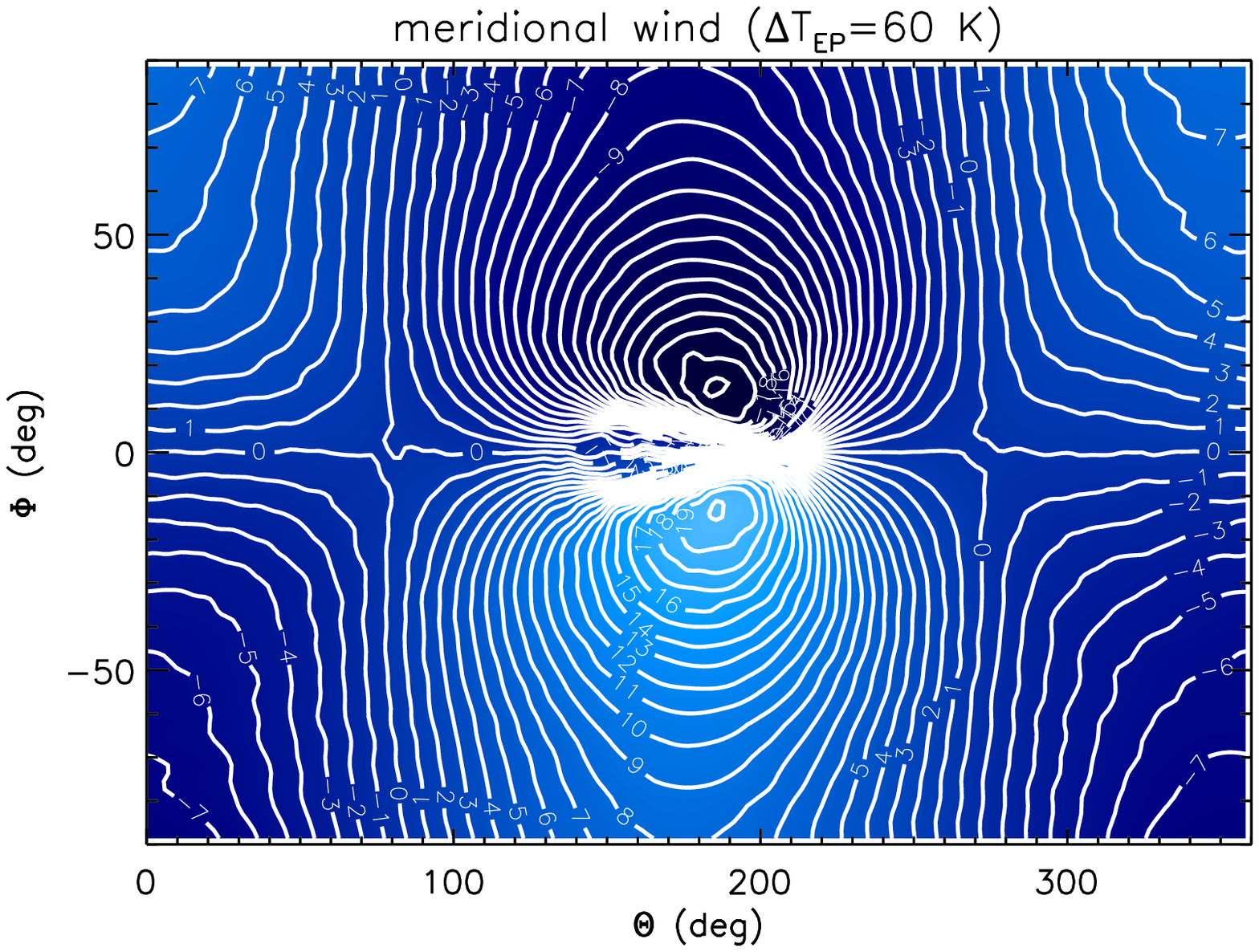}
\end{center}
\vspace{-0.2in}
\caption{Same as Figures \ref{fig:mass} and \ref{fig:rho}, but for three values of the equator-pole temperature difference: $\Delta T_{\rm EP}=20$ K (top row), $\Delta T_{\rm EP}=30$ K (middle row) and $\Delta T_{\rm EP}=60$ K (bottom row).}
\label{fig:dt}
\vspace{-0.15in}
\end{figure}

As described earlier, the qualitative structure of the global wind maps is insensitive to the adopted value of the equator-pole temperature difference $\Delta T_{\rm EP}$.  We now demonstrate explicitly that this is indeed the case.  Figure \ref{fig:dt} displays a set of global wind maps from simulations which assume tidal locking and for $\Delta T_{\rm EP}=20$ K, 30 K and 60 K.  While the qualitative structure of the wind maps are largely invariant to the variation of $\Delta T_{\rm EP}$, the quantitative predictions differ --- larger temperature differences between the equator and the poles lead to faster winds.  It is likely that the equator-pole temperature difference is lower on a tidally-locked exoplanet due to the reduced effect of Coriolis deflection, which is the reason why we have not tested models with greater values of $\Delta T_{\rm EP}$.  It is still possible that $\Delta T_{\rm EP} > 60$ K on exo-Earths in general, which will have strong consequences for the climate.

Nevertheless, the fact remains that $\Delta T_{\rm EP}$ is an unconstrained parameter in our models.  Until its value can be constrained by observations, \emph{quantitative} predictions for the wind speeds cannot be made \citep{tc10}.

\subsection{Varying the surface pressure $P_0$}
\label{subsect:p0}

\begin{figure}
\begin{center}
\includegraphics[width=0.48\columnwidth]{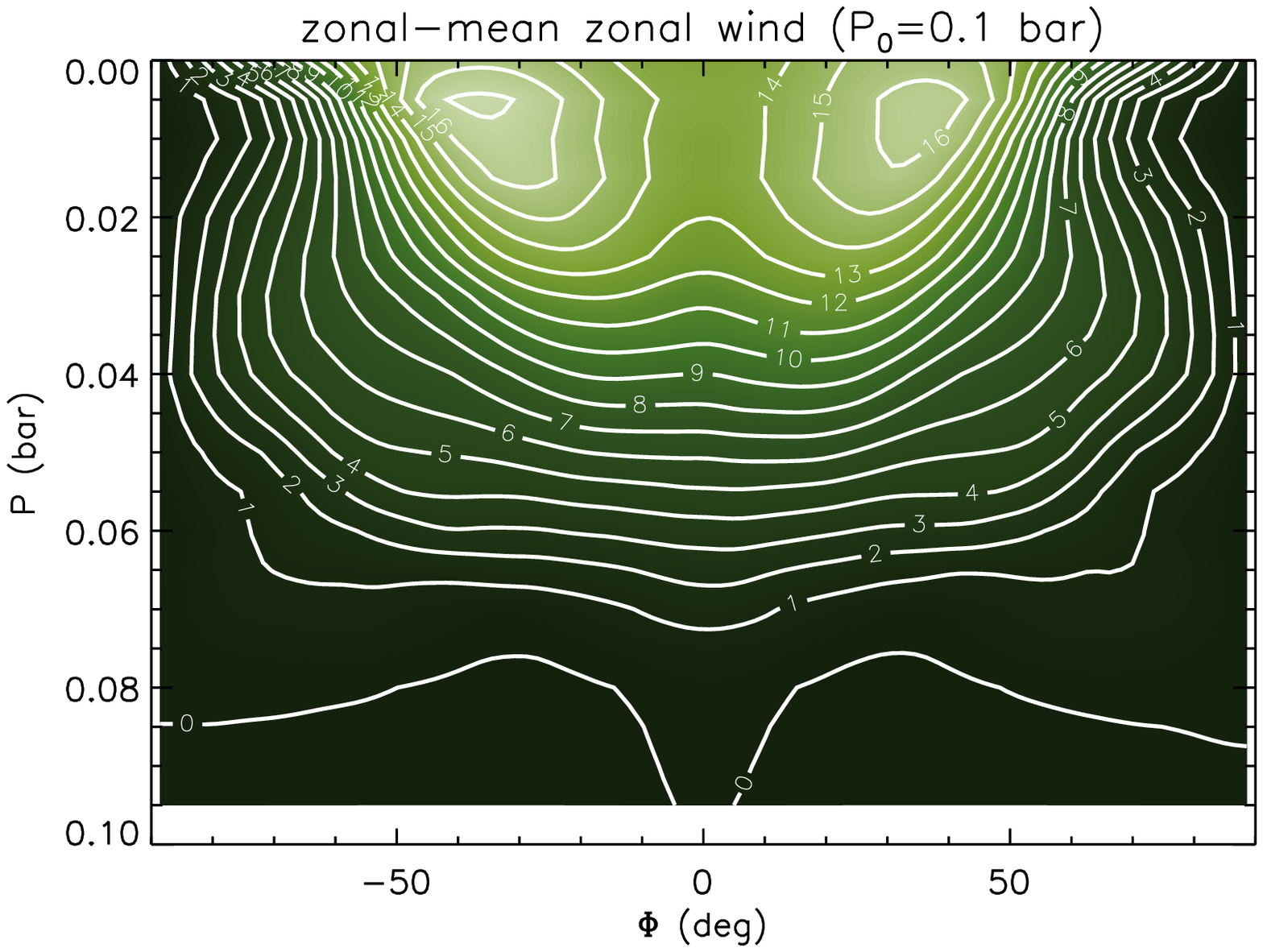}
\includegraphics[width=0.48\columnwidth]{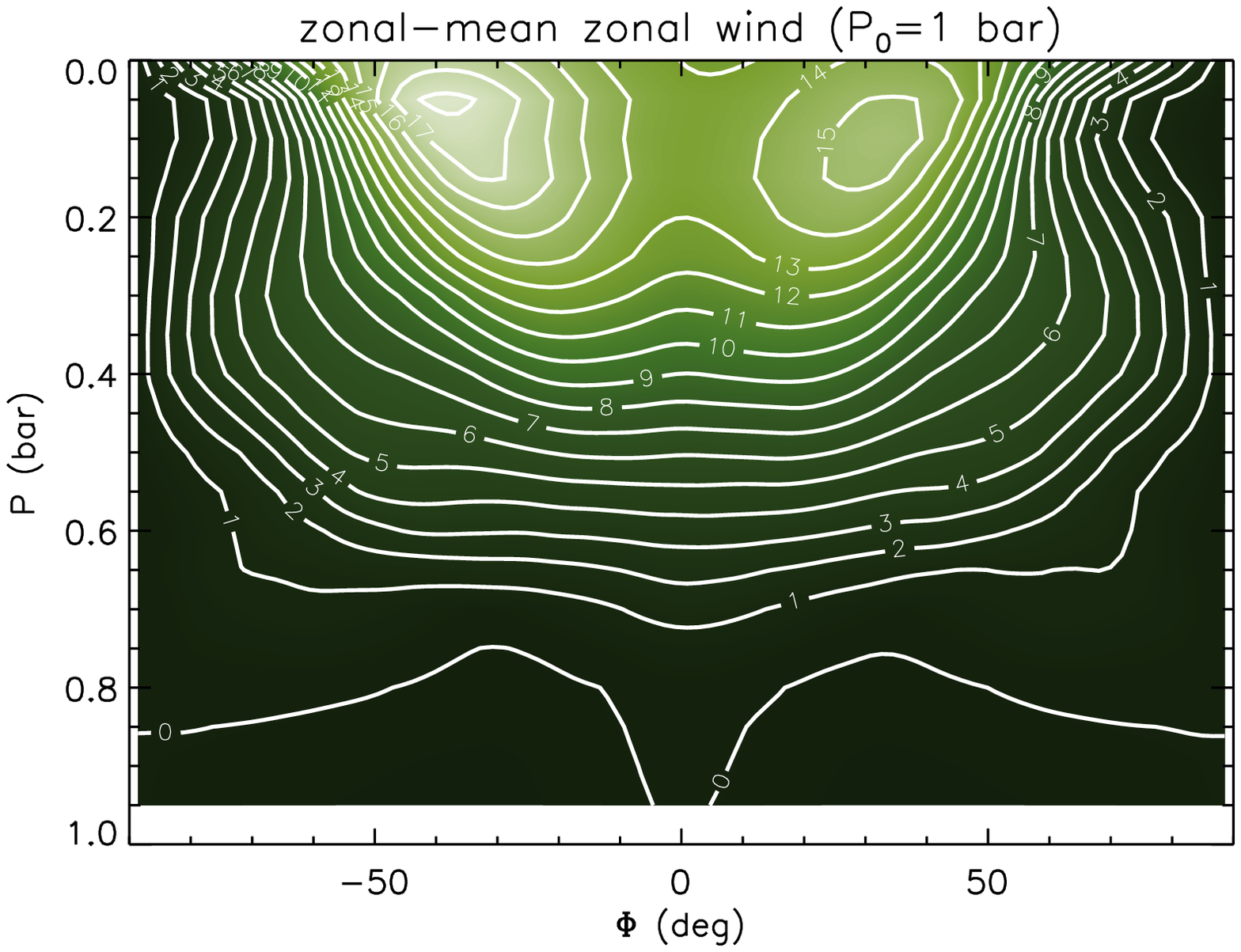}
\includegraphics[width=0.48\columnwidth]{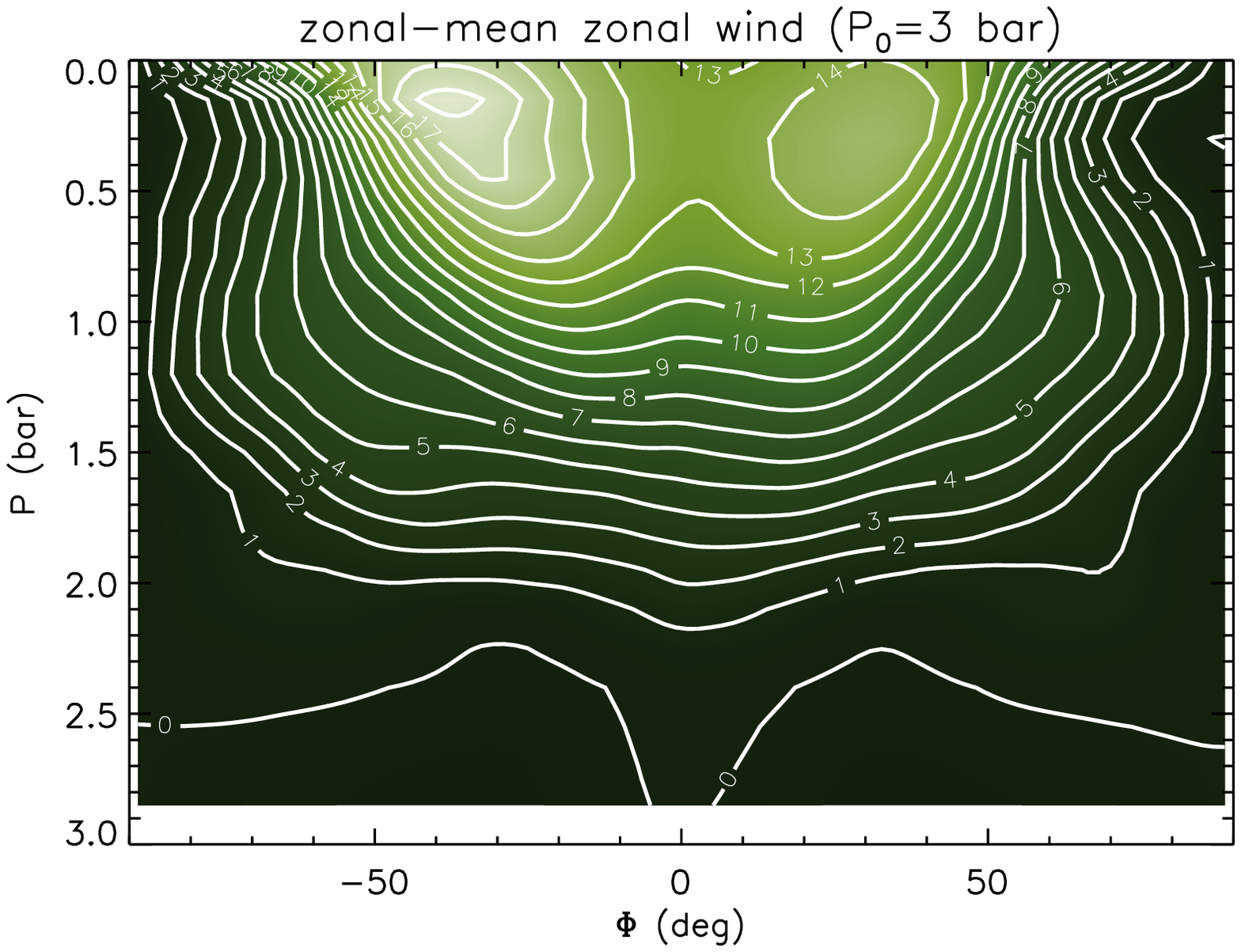}
\includegraphics[width=0.48\columnwidth]{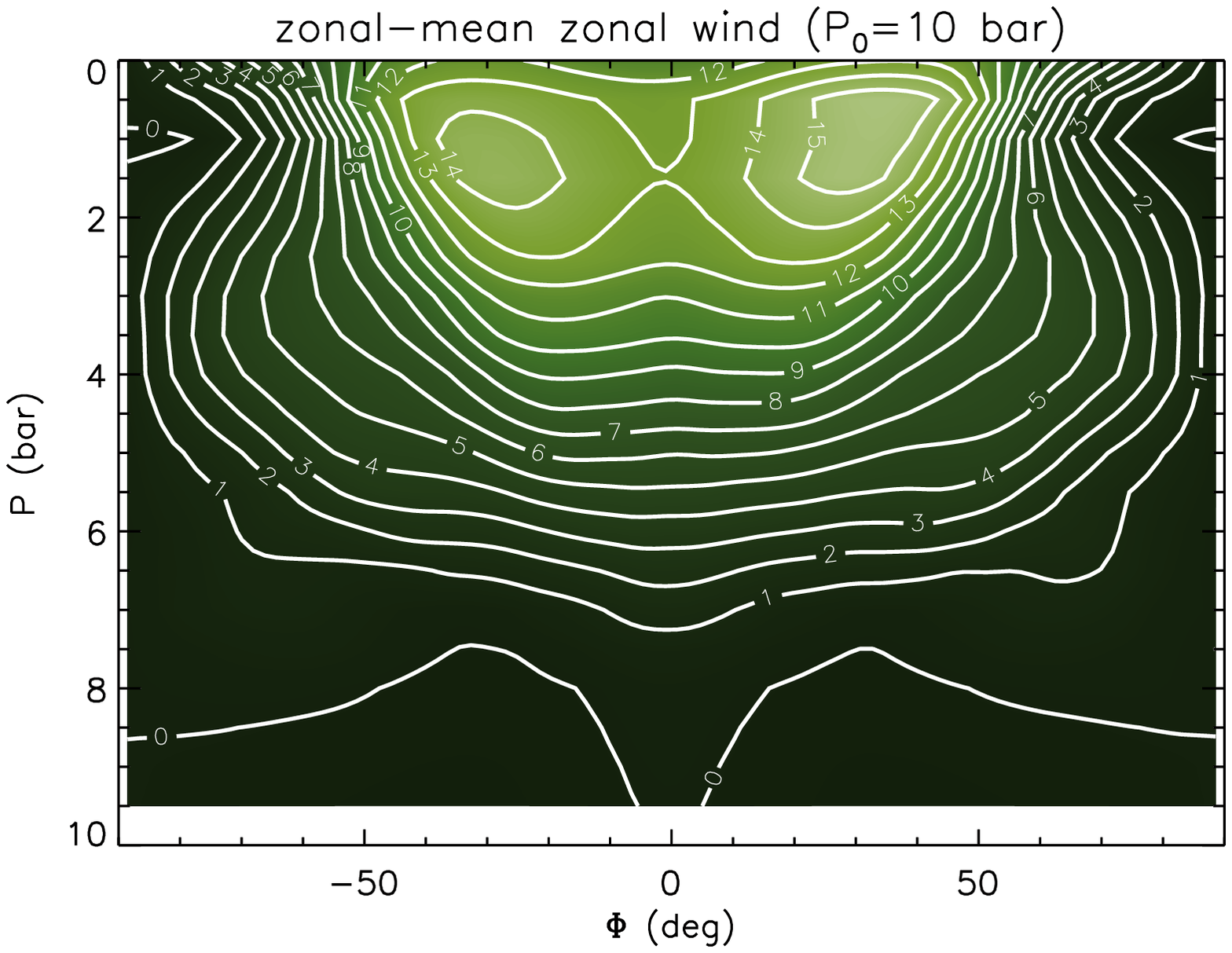}
\end{center}
\vspace{-0.2in}
\caption{Zonal-mean zonal wind profiles as functions of the vertical pressure $P$ and the latitude $\Phi$, where tidal locking is assumed.  Shown are simulations with different values of the vertical pressure: $P_0 = 0.1$ bar (top left panel) 1 bar (top right panel), 3 bar (bottom left panel) and 10 bar (bottom right panel).  Contour levels are given in m s$^{-1}$.}
\label{fig:p0}
\vspace{-0.1in}
\end{figure}

If Gliese 581g has an atmosphere, it may be thinner or thicker than that of Earth.  For comparison, Venus has a surface pressure $\sim 10^2$ bar due to its much thicker atmosphere, while the thin atmosphere of Mars produces $P_0 \sim 10^{-3}$ bar.  It is therefore conceivable that the surface pressure $P_0$ spans many orders of magnitude in value for exo-Earths.

To meaningfully compare simulations with different values of $P_0$ at the same vertical resolution (20 levels), we examine the zonal-mean zonal wind profiles as functions of the vertical pressure $P$ and latitude $\Phi$.  We term such statistically-averaged flow quantities the ``Held-Suarez statistics" \citep{hs94,hmp11}.  Figure \ref{fig:p0} shows results from simulations with $P_0 = 0.1$, 1, 3 and 10 bar.  For the range of surface pressures explored, we witness no qualitative differences in the zonal-mean zonal wind profiles, which are characterized by super-rotating, equatorial winds.  While we do not show them, we note that the global wind maps near the exoplanetary surface are very similar to those previously shown in Figures \ref{fig:mass} and \ref{fig:rho}.

While variations in the surface pressure appears to have little effect on the zonal-mean zonal wind profiles in the range of $P_0=0.1$--10 bar, it is possible that the actual surface pressure may be much less (thin atmosphere) or greater (thick atmosphere) than the range we have examined.

\subsection{The presence or absence of a tropopause}
\label{subsect:tropo}

\begin{figure}
\begin{center}
\includegraphics[width=0.48\columnwidth]{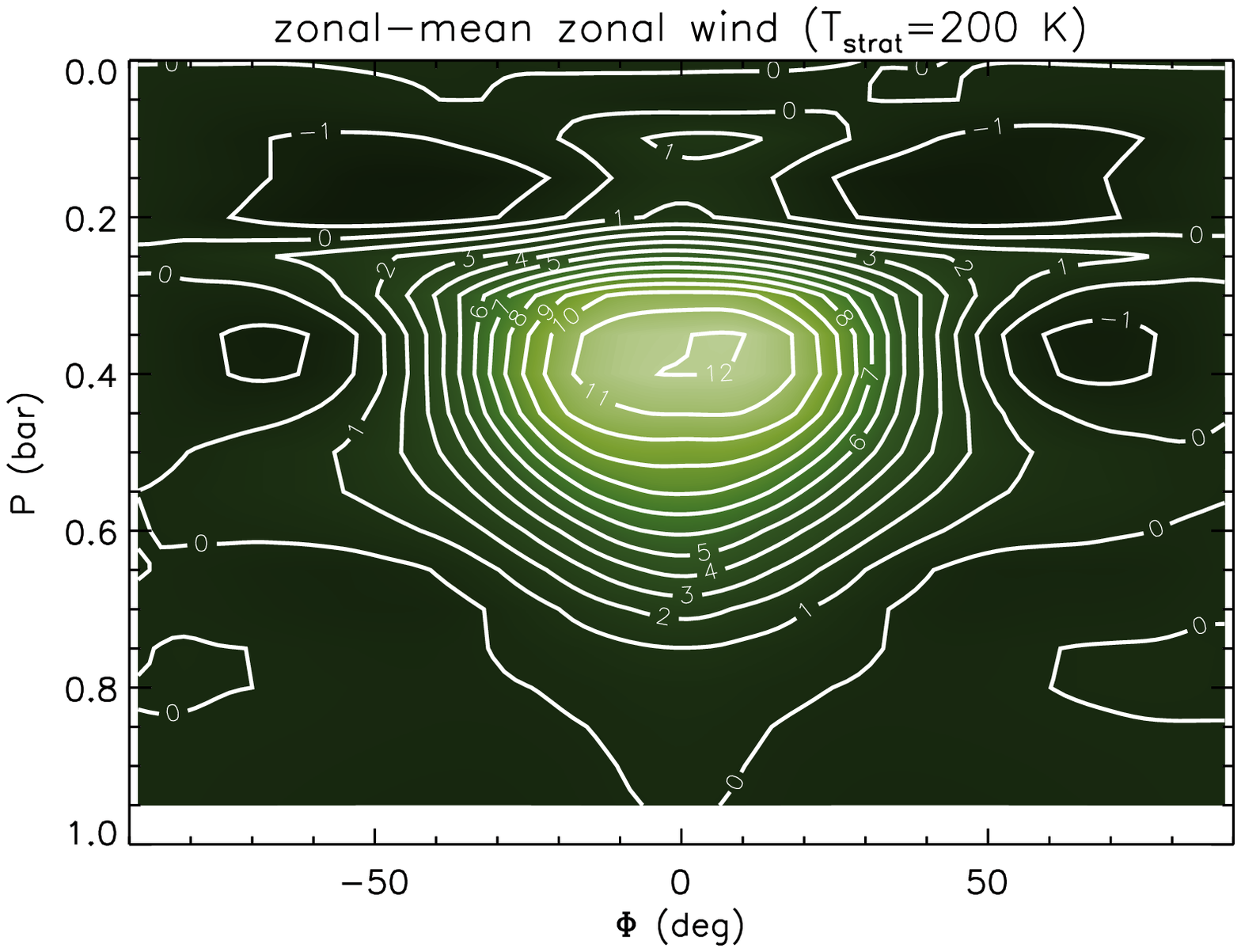}
\includegraphics[width=0.48\columnwidth]{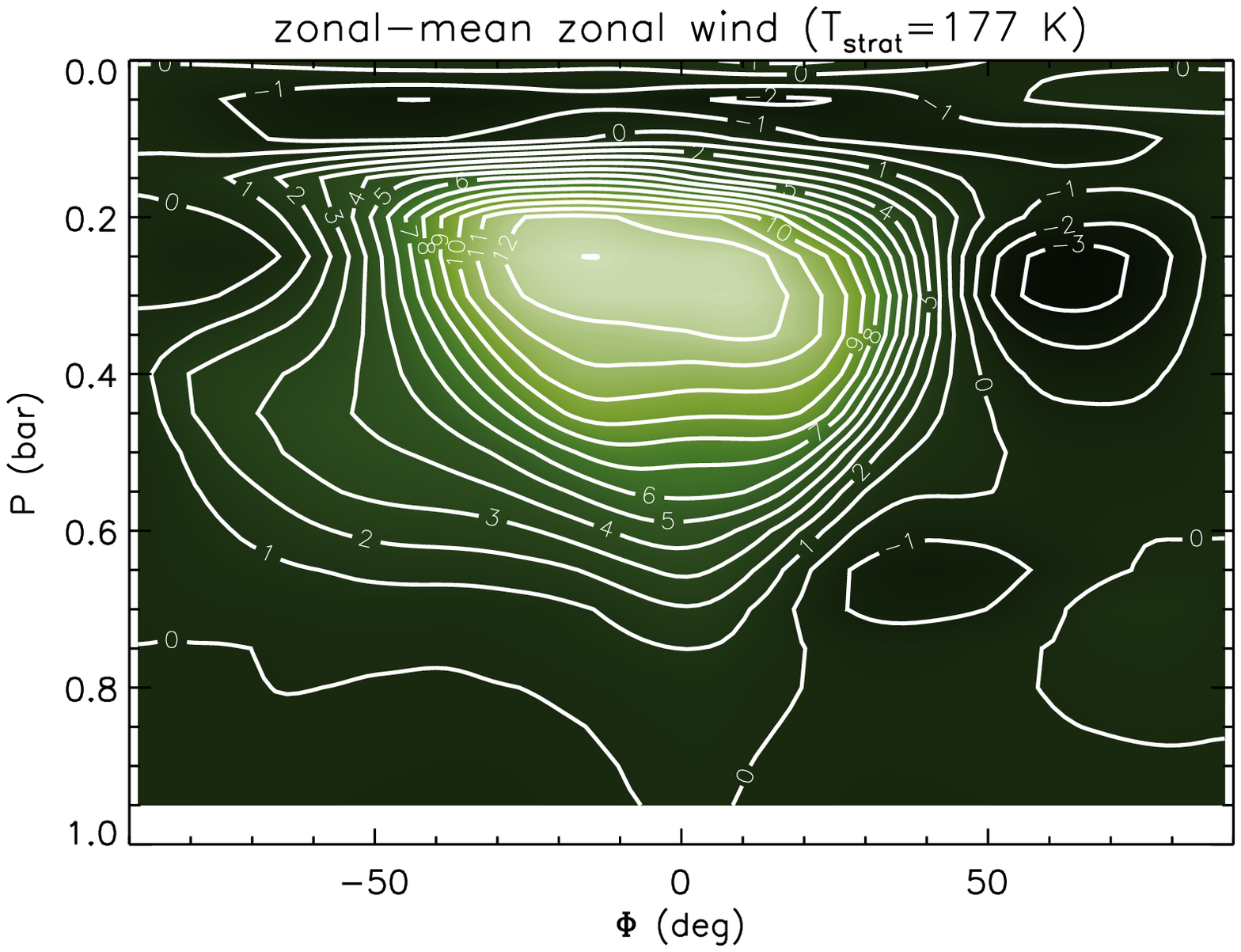}
\includegraphics[width=0.48\columnwidth]{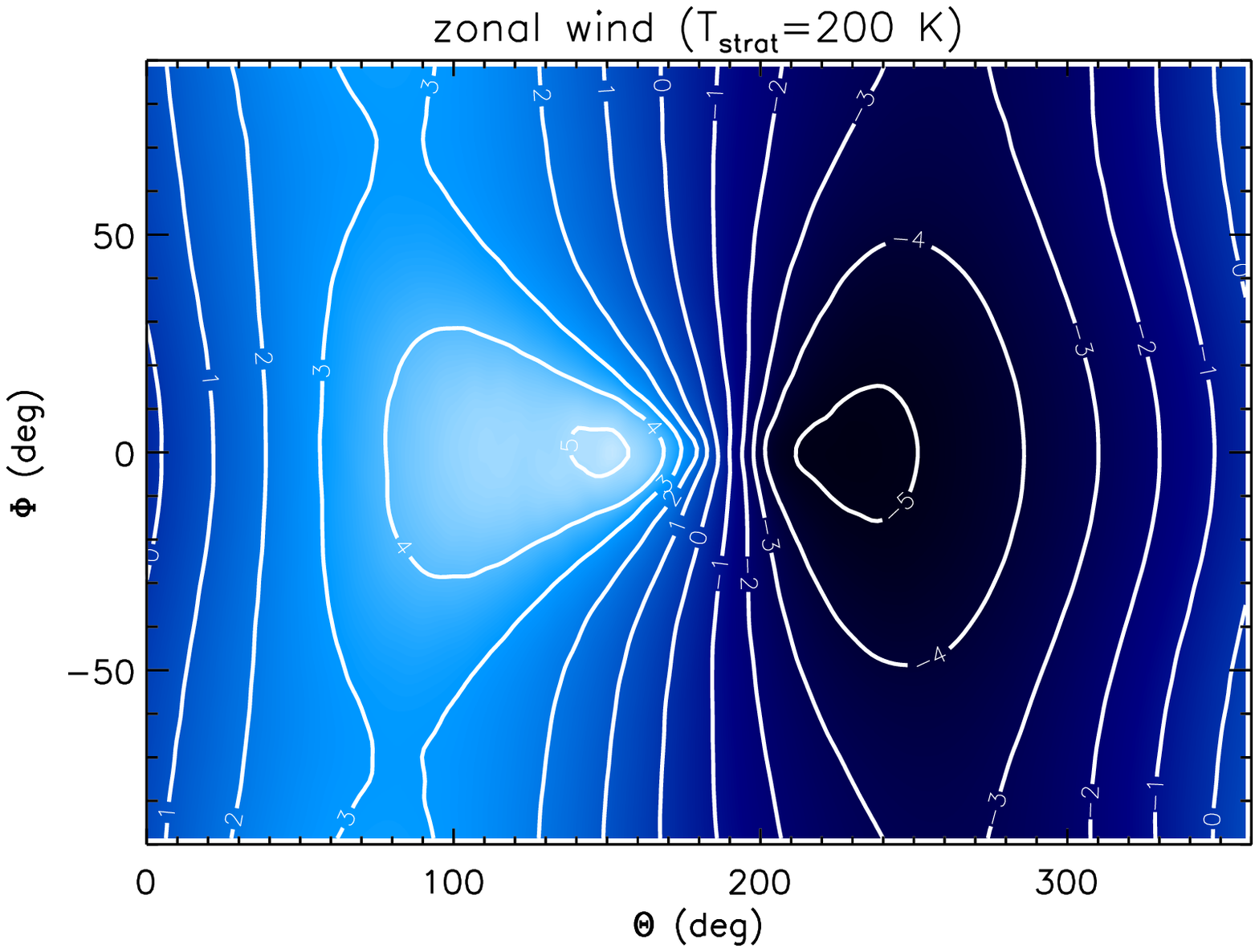}
\includegraphics[width=0.48\columnwidth]{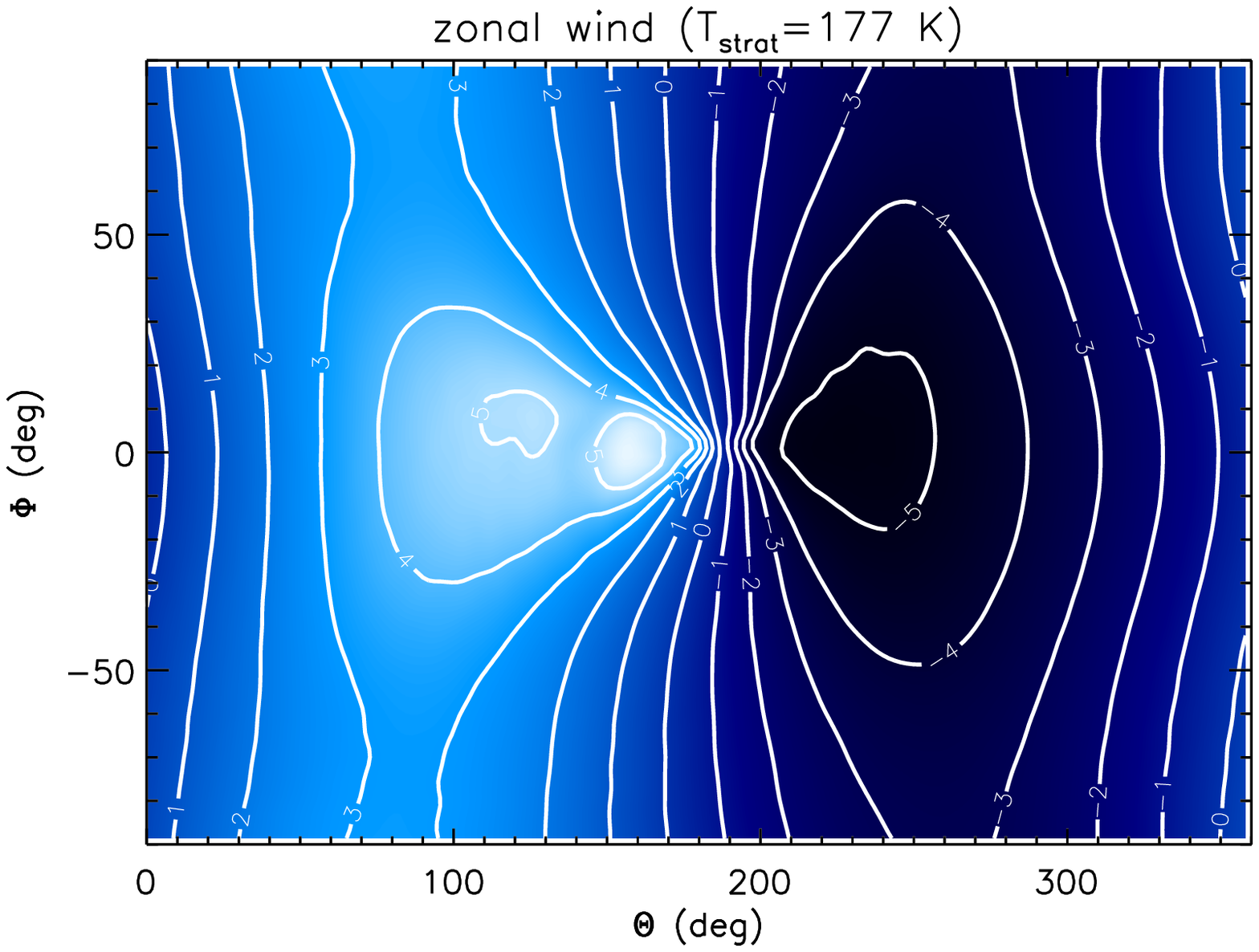}
\includegraphics[width=0.48\columnwidth]{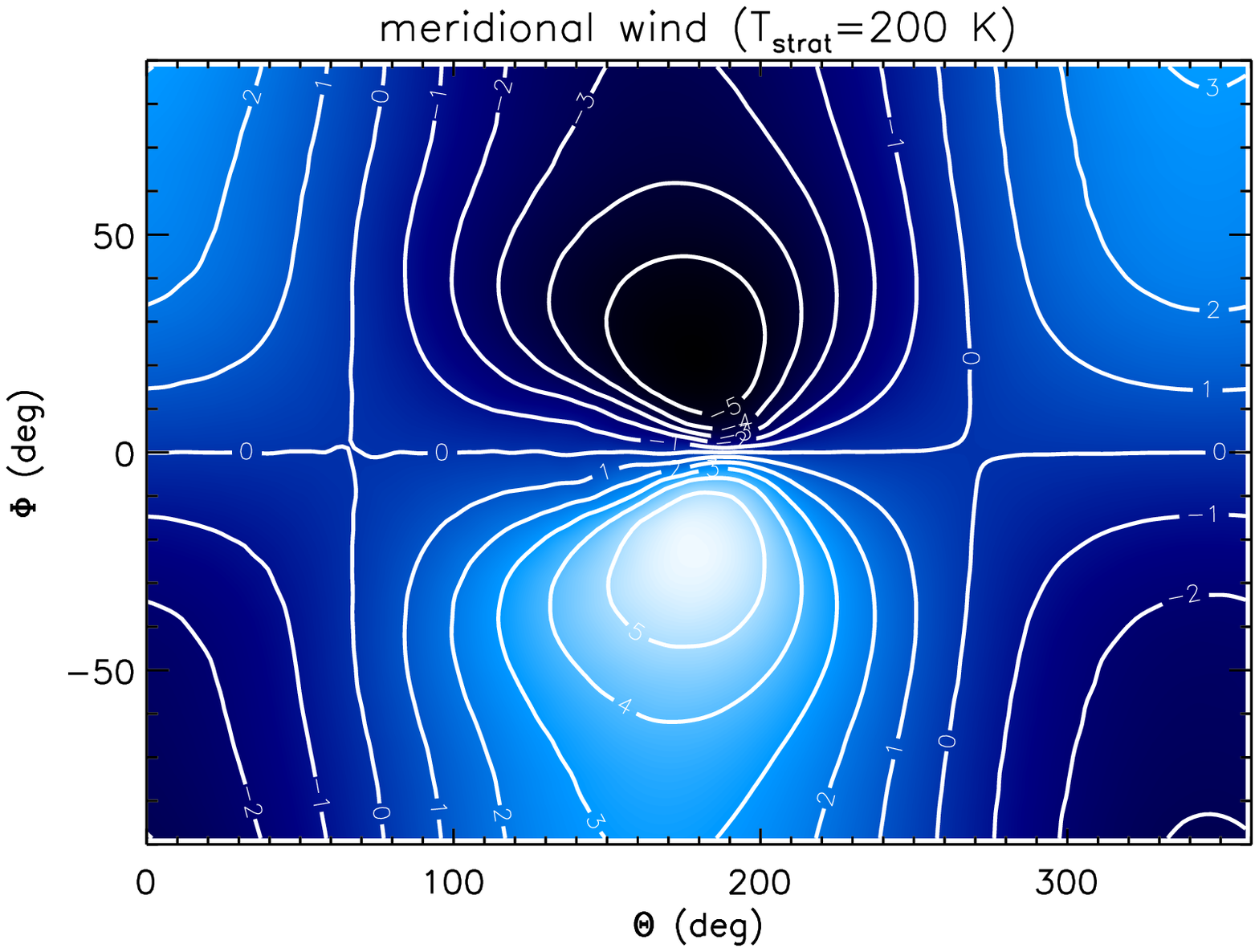}
\includegraphics[width=0.48\columnwidth]{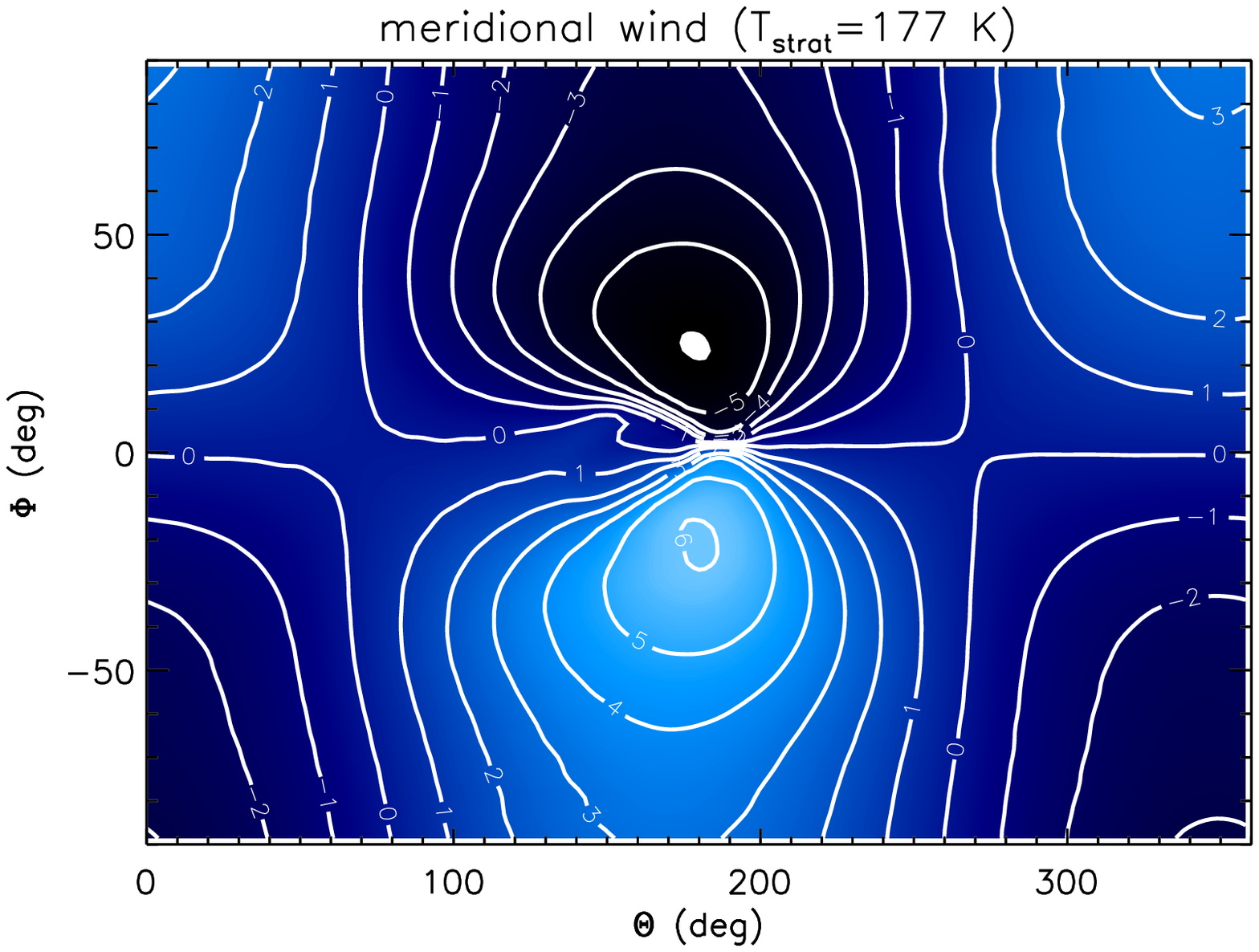}
\end{center}
\vspace{-0.2in}
\caption{Simulations including the presence of a tropopause, where the stratospheric temperature is denoted by $T_{\rm strat}$.  The top, middle and bottom panels show the zonal-mean zonal wind profiles, zonal wind maps for $P=0.95$--1 bar and meridional wind maps for $P=0.95$--1 bar, respectively.  The left and right columns are for simulations with $T_{\rm strat}=200$ K and 177 K, respectively (see text for details).  Contour levels are given in m s$^{-1}$.}
\label{fig:tropo}
\vspace{-0.1in}
\end{figure}

The terrestrial atmosphere closest to the surface may be divided into two components: the troposphere and the stratosphere.  The stratosphere is the barotropic component, meaning the surfaces of constant entropy are largely independent of latitude or longitude and increase with height.  By contrast, the surfaces of constant entropy change with latitude and longitude in the troposphere, making it the baroclinic component.  The height at which the troposphere transitions into the stratosphere is the tropopause.

To mimic the presence of a tropopause, \cite{hs94} implemented the following thermal forcing,
\begin{equation}
\tilde{T}_{\rm force} = \mbox{max}\left\{T_{\rm force}, T_{\rm strat} \right\},
\end{equation}
where $T_{\rm force}$ is given by equation (\ref{eq:ths}) and $T_{\rm strat} = 200$ K is the (constant) stratospheric temperature.  To further explore the variations on a theme associated with tidally-locked models, we use equation (\ref{eq:tidal}) for $T_{\rm force}$ and perform simulations with two different values of $T_{\rm strat}$.  The first retains $T_{\rm strat} = 200$ K as in the case of Earth, while the second uses equation (\ref{eq:scaling}) to scale the value of $T_{\rm strat}$ down to 177 K.  The point is not in the exact value of $T_{\rm strat}$ adopted but rather the change it causes in the zonal-mean zonal wind profile.

Figure \ref{fig:tropo} shows both the zonal-mean zonal wind profiles and the global wind maps produced by the simulations.  The zonal-mean zonal wind profile is characterized by a super-rotating, equatorial jet situated at $P \approx 0.2$--0.4 bar (for $P_0=1$ bar), where a lower stratospheric temperature corresponds to the jet residing higher up in the atmosphere simply because the tropopause is situated at a greater height in this case.  Interestingly, these differences do not translate into any qualitative differences for the global wind maps near the surface ($P=0.95$--1 bar) of the exoplanet, leading us to conclude that the presence of a tropopause does not have a major effect unless it reaches down to near the surface of the exoplanet.

\subsection{Varying the radiative cooling time $\tau_{\rm rad}$}
\label{subsect:tcool}

Our dynamical core simulations account for radiative cooling via the addition of an extra term to the thermodynamic equation \citep{hs94},
\begin{equation}
Q_{\rm Newton} = \left( T_{\rm force} - T \right) \left( \frac{1}{\tau_{\rm rad,d}} + {\cal Q} \right),
\end{equation}
where
\begin{equation} 
{\cal Q} \equiv 
\begin{cases}
0, & \sigma \le \sigma_b, \\
\left(\frac{1}{\tau_{\rm rad,u}} - \frac{1}{\tau_{\rm rad,d}} \right) \left(\frac{\sigma-\sigma_b}{1-\sigma_b}\right) \cos^4\Phi, & \sigma > \sigma_b, \\
\end{cases}
\end{equation}
the dimensionless pressure is $\sigma \equiv P/P_0$ and the quantity $\sigma_b$ is set to 0.7 following \cite{hs94}.  In all of the simulations presented so far, we have kept $\tau_{\rm rad,u} = 4$ Earth days and $\tau_{\rm rad,d} = 40$ Earth days.  In this subsection, we explore the implications of increasing both $\tau_{\rm rad,u}$ and $\tau_{\rm rad,d}$ by a constant, multiplicative factor.  For convenience, we refer to the fiducial radiative cooling time simply as $\tau_{\rm rad} = \tau_{\rm rad,u} = 4$ Earth days, where it is implied that $\tau_{\rm rad,d}/\tau_{\rm rad,u} = 10$.  The damping of low-level winds due to the drag or friction between the atmosphere and exoplanetary surface is treated by applying the Rayleigh damping coefficient,
\begin{equation}
{\cal D}_{\rm Rayleigh} =
\begin{cases}
0, & \sigma \le \sigma_b, \\
\frac{\sigma-\sigma_b}{\tau_{\rm fric} \left(1-\sigma_b\right)}, & \sigma > \sigma_b, \\
\end{cases}
\end{equation}
to the velocity field.  The fiducial value of the drag time is $\tau_{\rm fric} = 1$ Earth day.  When we increase the radiative cooling time, we apply the same constant, multiplicative factor to the Rayleigh drag time.

We have chosen the Held-Suarez benchmark as a baseline because of its simplicity.  While there is no strict justification for extending the Newtonian cooling and Rayleigh drag schemes to exo-Earths, these are also the simplest schemes one can adopt to mimick the effects of cooling and drag.

\begin{figure}
\begin{center}
\includegraphics[width=0.48\columnwidth]{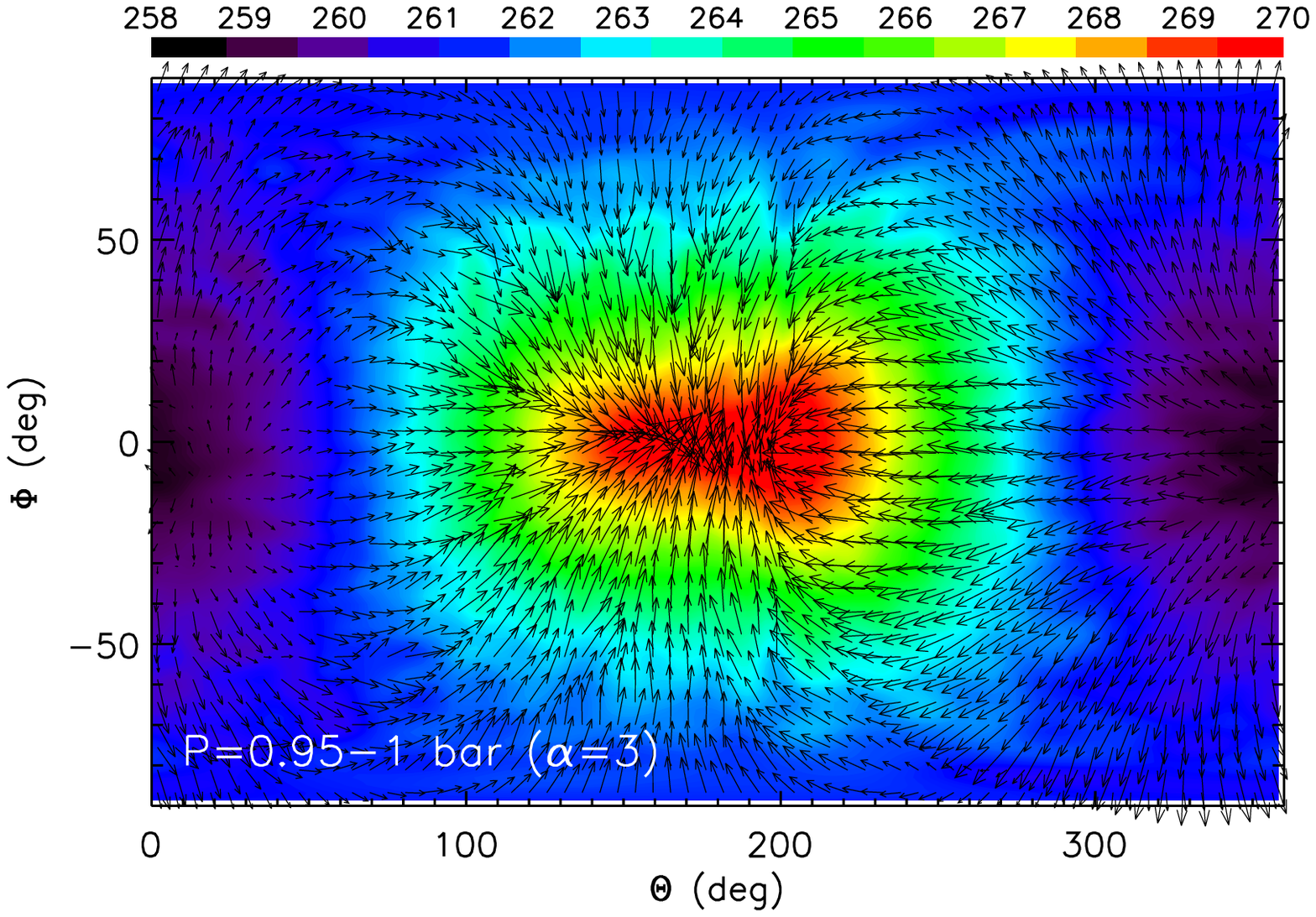}
\includegraphics[width=0.48\columnwidth]{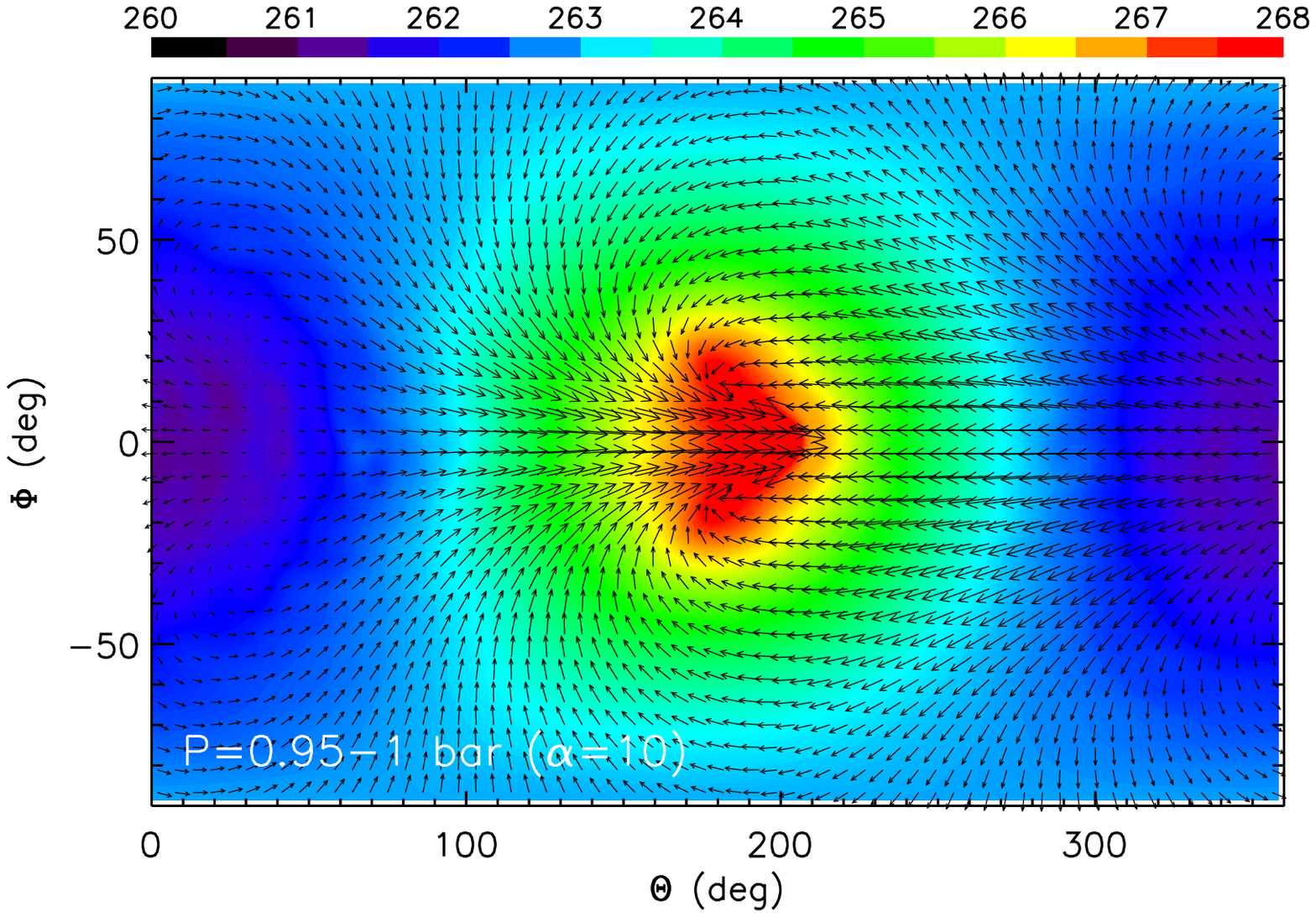}
\includegraphics[width=0.48\columnwidth]{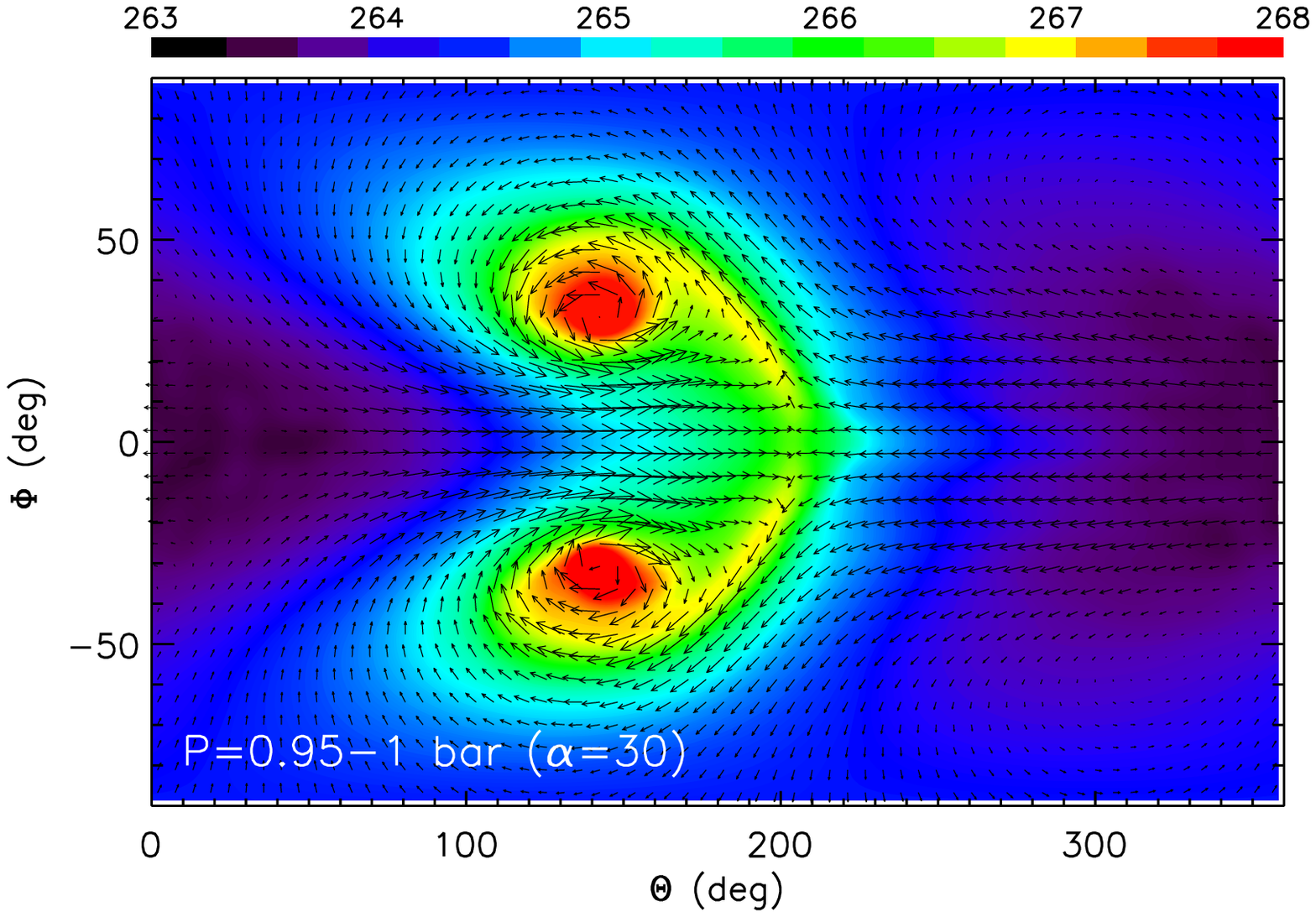}
\includegraphics[width=0.48\columnwidth]{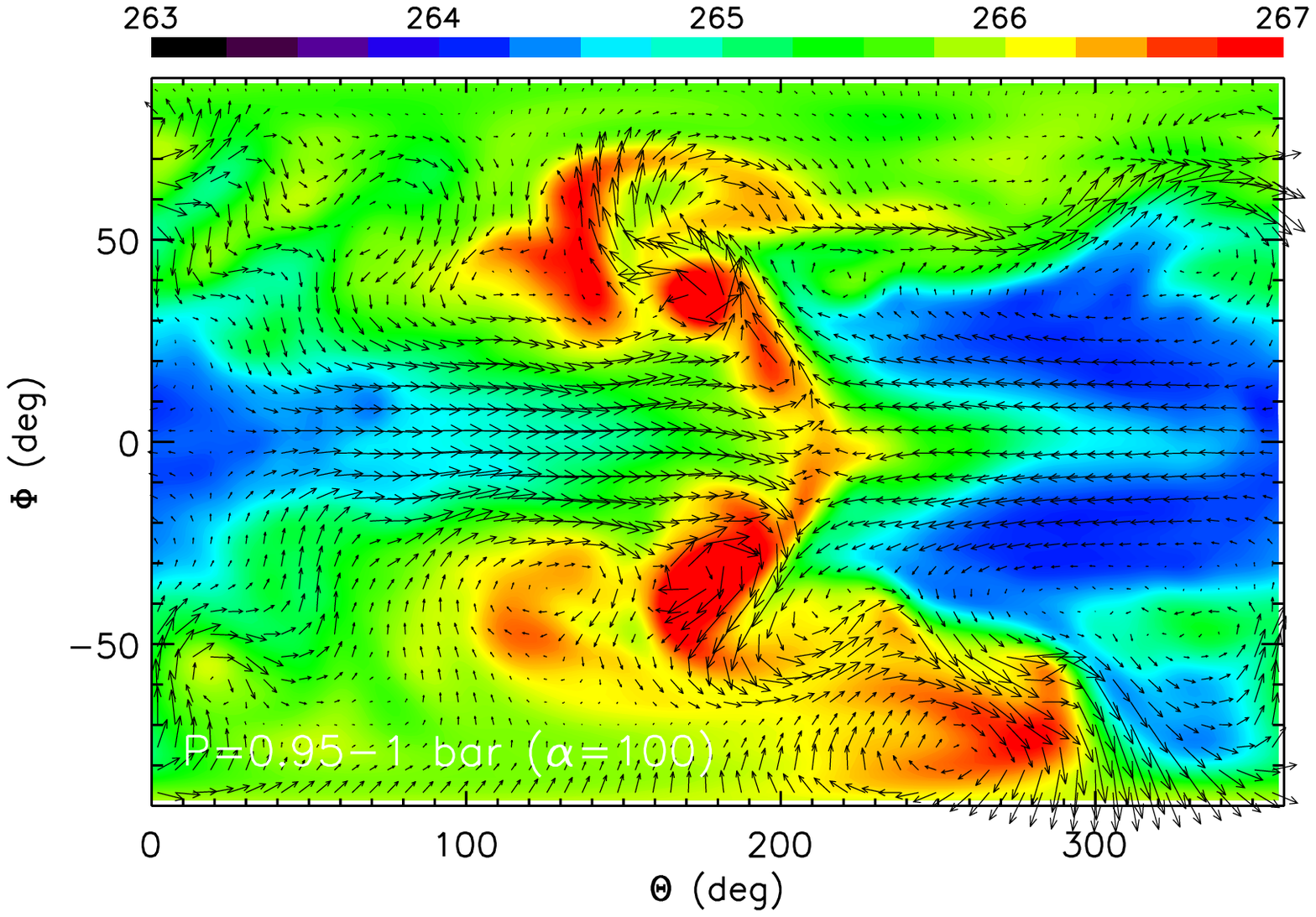}
\end{center}
\vspace{-0.2in}
\caption{Snapshots of the temperature and velocity fields near the surface ($P=0.95$--1 bar), taken from simulations where the radiative cooling and Rayleigh friction time scales have been increased by factors of $\alpha = 3$ (top left panel), 10 (top right panel), 30 (bottom left panel) and 100 (bottom right panel).  Colors denote temperature (in K) and the arrows represent the direction of the velocity field.  The resolution adopted is T63L20 ($192 \times 96 \times 20$).  The snapshots are taken at 3000 Earth days after the start of the simulations.}
\label{fig:cool2}
\vspace{-0.1in}
\end{figure}

We now explore models where $\tau_{\rm rad}$ and $\tau_{\rm fric}$ are increased by factors of $\alpha = 3$, 10, 30 and 100 while retaining the assumption of tidal locking.  Figure \ref{fig:cool2} shows snapshots of the temperature and velocity fields from simulations that have attained quasi-equilibrium.  As $\alpha$ increases and radiative cooling becomes less efficient, advection becomes more effective at transporting parts of the atmosphere heated by stellar irradiation from the day to the night side.  Thus, the simple picture of segregated hemispheres where one hemisphere is the permanent day side and the other is the permanent night side starts to break down, at least from the view point of a temperature map.  Some of these models are refutable, because when astronomical instrumentation becomes advanced enough to perform direct wind measurements of exo-Earths --- such as was done for the hot Jupiter HD 209458b by \cite{snellen10}, who measured the CO line in absorption with the line center blueshifted by about 2 km s$^{-1}$ --- the predictions for the recorded spectrum will be different for each of these scenarios.  For example, if advection dominates over radiative cooling and succeeds in the longitudinal homogenization of the temperature field, then the spectrum will have both blue- and redshifted contributions to the absorption lines obtained from transmission spectroscopy, thus cancelling out the spectral shift from line center.

It remains to be determined, from both theory and observations, if the radiative cooling and Rayleigh friction times assumed in our models are consistent with the atmospheric chemistry actually found on exo-Earths.

\subsection{Varying the rotational frequency $\Omega_p$}
\label{subsect:rotate}

\begin{figure}
\begin{center}
\includegraphics[width=0.48\columnwidth]{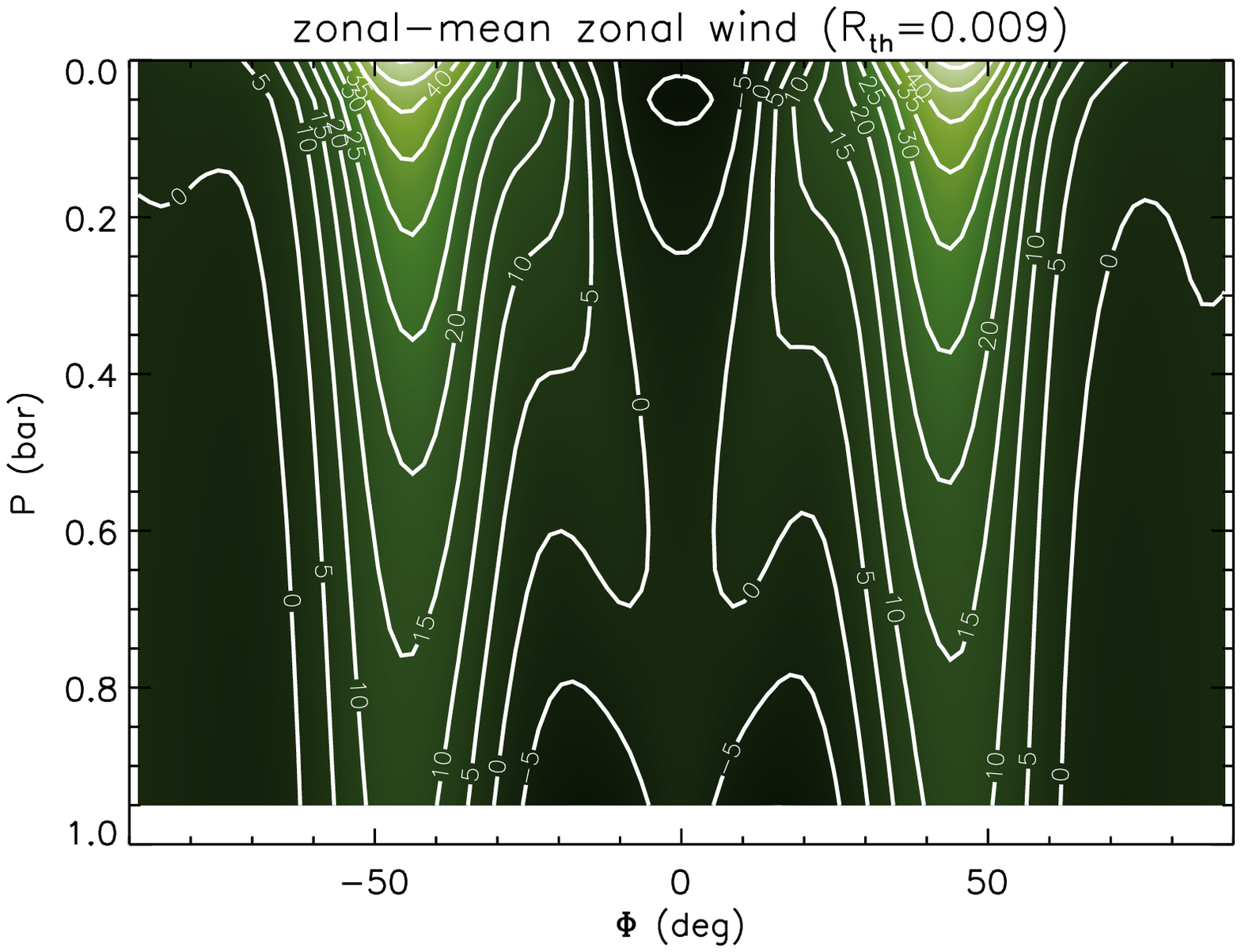}
\includegraphics[width=0.48\columnwidth]{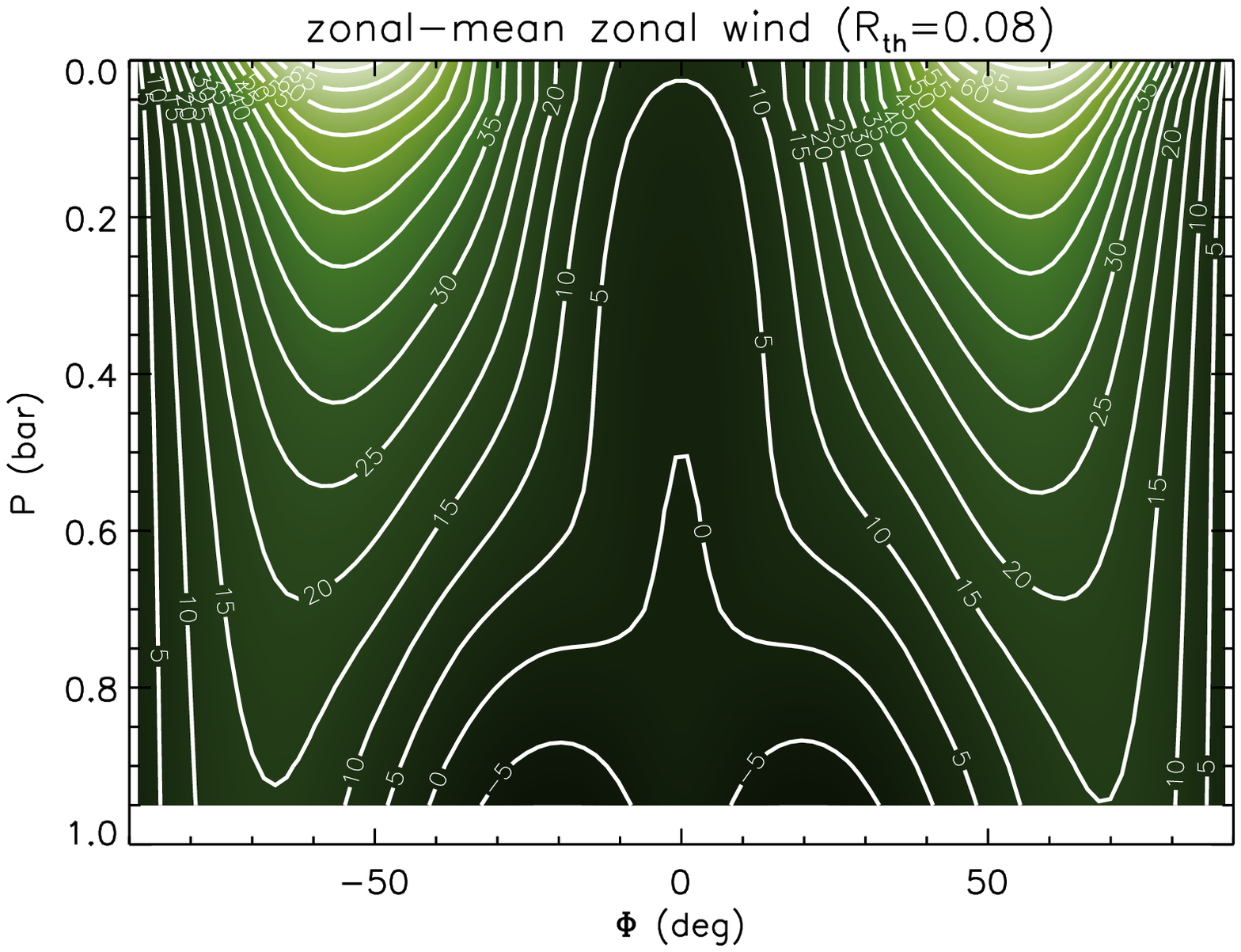}
\includegraphics[width=0.48\columnwidth]{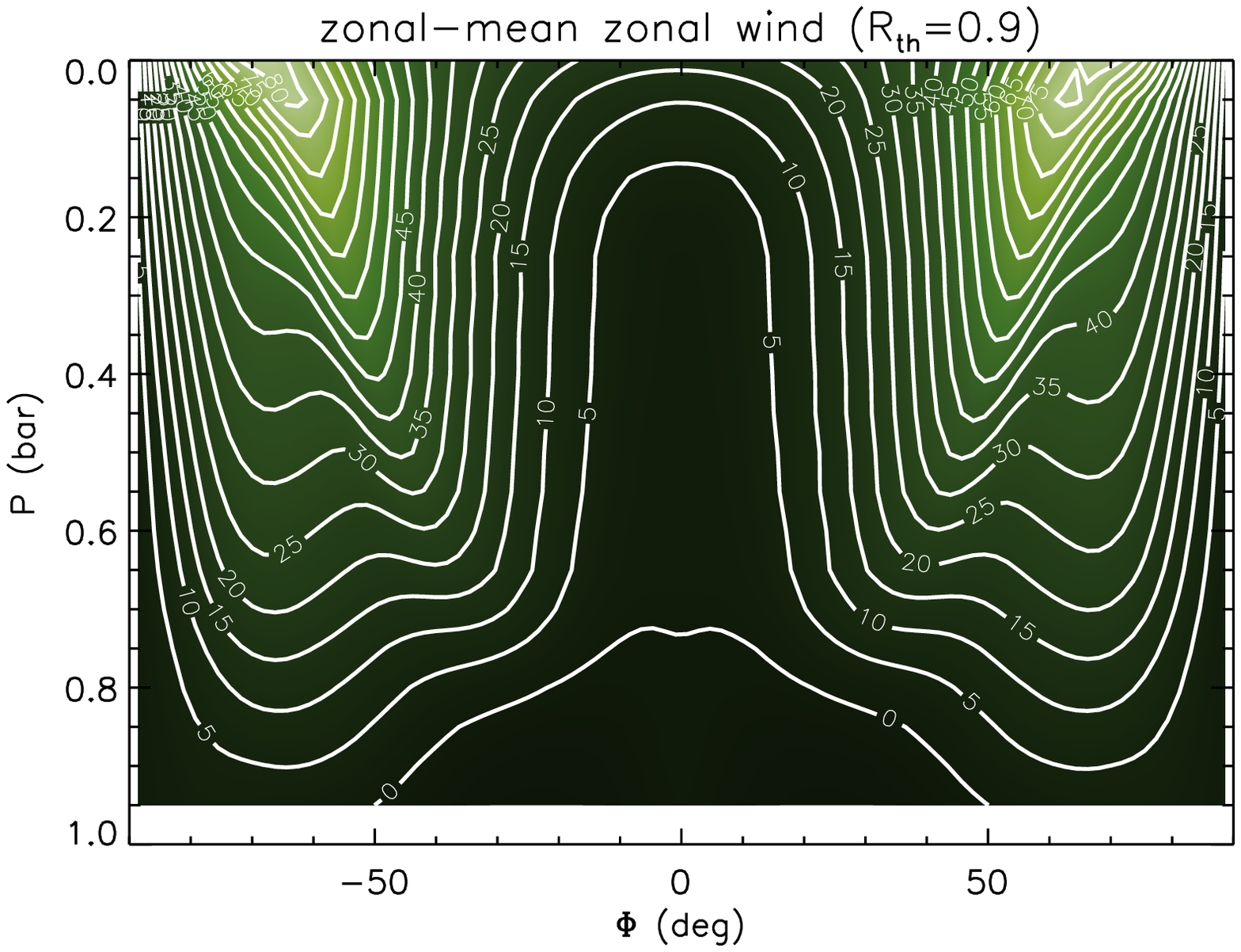}
\includegraphics[width=0.48\columnwidth]{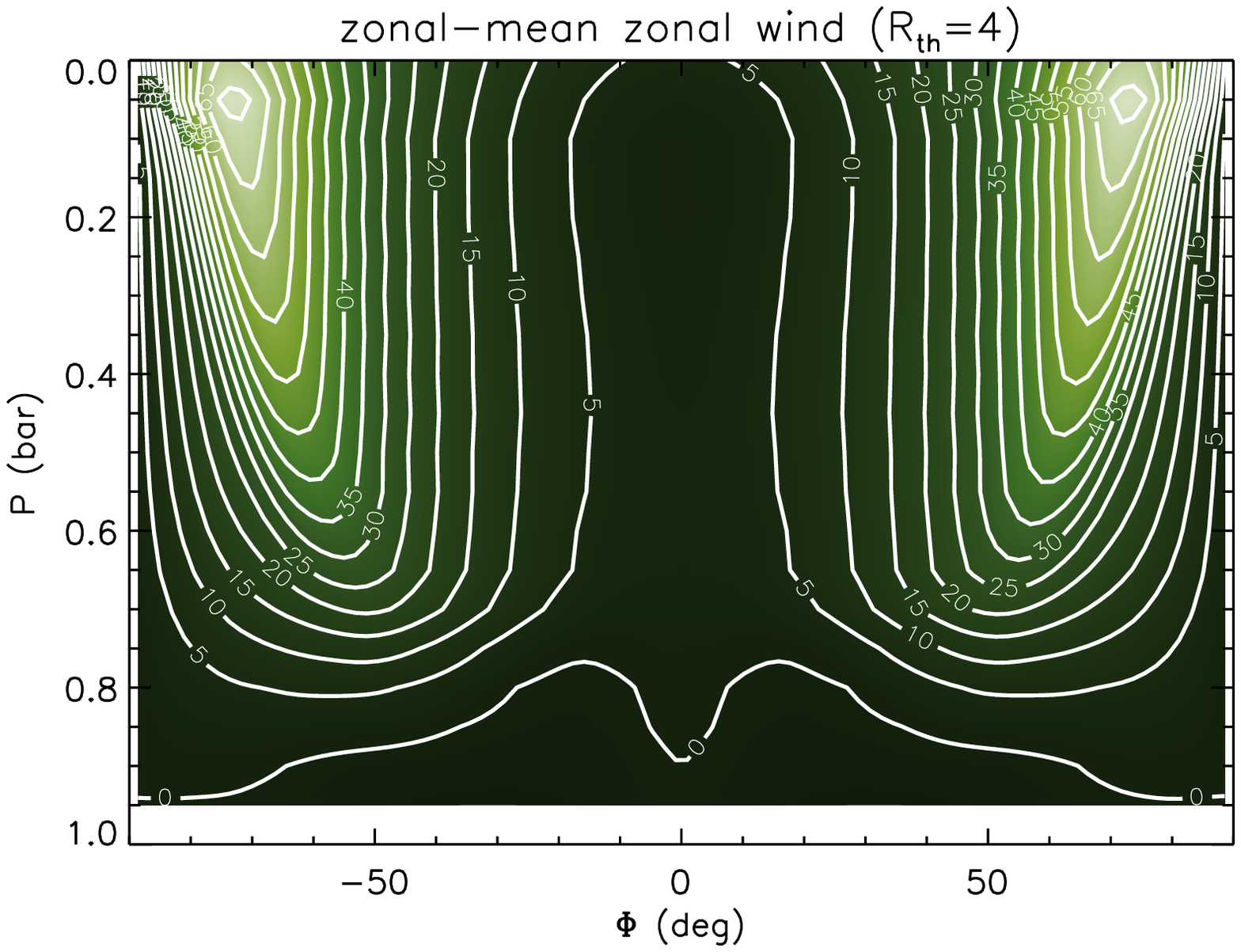}
\end{center}
\vspace{-0.2in}
\caption{Zonal-mean zonal wind profiles as functions of the vertical pressure $P$ and the latitude $\Phi$, where the assumption of tidal locking is relaxed.  Shown are simulations with different values of the thermal Rossby number: ${\cal R}_{\rm th} \approx 0.009$ (1 Earth day rotation; top left panel), ${\cal R}_{\rm th} \approx 0.08$ (3 Earth day rotation; top right panel), ${\cal R}_{\rm th} \approx 0.9$ (10 Earth day rotation; bottom left panel) and ${\cal R}_{\rm th} \approx 4$ (20 Earth day rotation; bottom right panel).  Contour levels are given in m s$^{-1}$.}
\label{fig:omega}
\vspace{-0.1in}
\end{figure}

The axisymmetric thermal forcing described in equation (\ref{eq:ths}) is valid when the assumed rotational period is less than the orbital period, e.g., in the second baseline model presented in \S\ref{subsect:baseline} with $\Omega_p = 7.292 \times 10^{-5}$ s$^{-1}$ (one Earth day).  As the last variation on a theme, we explore models where the assumption of tidal locking is removed and the rotational period is assumed to be 1, 3, 10 and 20 Earth days (and where $\tau_{\rm rad}$ and $\tau_{\rm fric}$ have been reverted to their fiducial values).  Since the observed orbital period of Gliese 581g is about 37 days, we can use equation (\ref{eq:ths}) for the thermal forcing function.  Such a parameter exploration of the \emph{rotational} (and not orbital) frequency $\Omega_p$ is not unreasonable because tidal locking is a theoretical expectation based on the extrapolation of the properties of Solar System planets --- which may not be representative of the exoplanets in the Universe at large --- and not a direct astronomical observable.

In this regime of axisymmetric thermal forcing, exoplanets may be characterized by the thermal Rossby number \citep{mv10},
\begin{equation}
{\cal R}_{\rm th} = \frac{ {\cal R} ~\Delta T_{\rm EP} }{\left( 2 \Omega_p R_p \right)^2} \approx 9 \times 10^{-3} ~\left( \frac{{\cal R}}{287.04 \mbox{ J kg}^{-1} \mbox{ K}^{-1}} \frac{\Delta T_{\rm EP}}{60 \mbox{ K}} \right)  \left( \frac{\Omega_p}{7.292 \times 10^{-5} \mbox{ s}^{-1}} \frac{R_p}{9290 \mbox{ km}} \right)^{-2},
\end{equation}
where ${\cal R}$ is again the ideal gas constant.  Since ${\cal R}_{\rm th} < 1$, a hypothetical Gliese 581g with a rotational period of 1 Earth day has a counter-rotating, equatorial wind, as shown in the right panel of Figure \ref{fig:zonal} and also Figure \ref{fig:wind}.  When the rotational period is increased to 3, 10 and 20 Earth days, the thermal Rossby number becomes ${\cal R}_{\rm th} \approx 0.08$, 0.9 and 4 respectively.  In Figure \ref{fig:omega}, we clearly see a transition from equatorial counter-rotation to the beginnings of super-rotation (near the surface; $P \approx 0.8$--1 bar) as ${\cal R}_{\rm th}$ is increased --- equatorial super-rotation is expected to occur when ${\cal R}_{\rm th} \gg 1$ \citep{mv10}.  However, even with a rotational period of 20 Earth days (bottom right panel of Figure \ref{fig:omega}), we are already approaching the limit where the axisymmetric thermal forcing in equation (\ref{eq:ths}) starts to become invalid and therefore we do not perform more simulations with even longer rotational periods.  Rather, our results in Figure \ref{fig:omega} are already sufficient to conclude that the rotational frequency $\Omega_p$ is a major parameter of the system.

Finally, we note that equatorial super-rotation has previously been studied within the context of non-axisymmetric thermal forcing \citep{sd92,sar93,kh05,sp10}.  It has also been previously studied using a more direct, non-axisymmetric \emph{momentum} forcing \citep{sh04,cho08}.

\section{Discussion}
\label{sect:discussion}

We have presented three-dimensional simulations of atmospheric circulation using the exo-Earth candidate Gliese 581g as a test bed.  Our starting point is a dynamical model that is similar to the Held-Suarez benchmark for Earth, which is calibrated to reproduce terrestrial observations of large-scale climate patterns.  Using this model as a baseline, we then present global temperature and wind maps which assume Gliese 581g to be tidally locked.  \emph{The salient, qualitative conclusion gleaned from our study is that a ``dance" occurs between the exoplanet and its host star.}  (See also \citealt{cho08}.)  When radiative cooling dominates over advection, the star successfully imprints its thermal signature on the exoplanet, which then resembles a sphere painted ``half black and half white" (left panels of Figures \ref{fig:mollweide} and \ref{fig:temperature}) akin to the way one usually visualizes the permanent day and night sides of a tidally-locked exoplanet in a cartoon or schematic diagram.  When advection occurs more efficiently than radiative cooling, the exoplanet leads the dance and smears out the thermal forcing imposed by the star evenly across longitude (right panel of Figure \ref{fig:temperature}).  In the intermediate regime between these two cases, the exoplanet only partially succeeds in the longitudinal homogenization of temperature, an example of which is shown in the right panel of Figure \ref{fig:mollweide}.  In this case, the simple picture of the permanent day and night sides becomes more involved.

In the context of the classical $0^\circ$--$100^\circ$C range of temperatures for habitability, our main finding is that the specific locations for habitability on the surface of Gliese 581g --- and exo-Earths in general --- depend on whether the exoplanet is tidally-locked and how fast radiative cooling occurs on a global scale.  A shortcoming of our approach is that we have neglected the effects of radiative transfer, which may have little effect on the large-scale climate patterns but will almost certainly alter the absolute values of the temperatures --- for example, including the effects of CO$_2$ may result in temperatures warmer than what we find (i.e., the greenhouse effect).  The age-old question of whether an exoplanet is inhabitable constitutes an active field and a broad range of approaches has been adopted by various researchers.  \cite{sel07} used 1D radiative-convective atmospheric models to assess the conditions on Gliese 581c and 581d, and find the latter to be potentially the first habitable exoplanet discovered.  \cite{word10} also used a 1D radiative-convective scheme, which included CO$_2$ and H$_2$O clouds, to examine the habitability of Gliese 581d.  They find CO$_2$ to be an important ingredient in maintaining global temperatures above the freezing point of water --- increasing the surface gravity of the exoplanet reduces the CO$_2$ column density, which results in global cooling of its surface.  They also find Rayleigh scattering to be less important for the redder spectra of M class stars.  \cite{word10b} followed up with preliminary results from a three-dimensional climate model that focused on a pure CO$_2$ atmosphere, using the early Martian atmosphere as a baseline.  They raised the concern that CO$_2$ is able to condense out in the colder regions of the exoplanet and may lead to (CO$_2$) atmospheric collapse.  \cite{dvo10} examine the interplay between the dynamical stability of systems with multiple stars and/or exoplanets and their habitable zones.  \cite{lammer10} emphasize the coupling between the geophysical activity of the exoplanet and the properties of its parent star.  Even the definition of the inner edge of the habitable zone depends on the amount of cloud cover on the exoplanet \citep{sel07}.  Generally, these studies make the point that an exoplanet found outside of the classical habitable zone may not be uninhabitable --- conversely, an exoplanet found within the zone may not be inhabitable.  Furthermore, a seemingly inhospitable exoplanetary surface does not exclude the possibility of subsurface habitability \citep{gold92}.

Although Gliese 581 itself is observed to exhibit weak Ca~{\sc ii} H and K emission \citep{mayor09}, it is well known that M dwarfs in general are chromospherically active \citep{sh86,scalo07} and it is likely that stellar activity will modify the exoplanetary atmosphere away from the terrestrial baseline \citep{sel07}.  It is also plausible that slower rotation results in weaker magnetospheres, which in turn weakens the ability of the exoplanet to shield itself from stellar irradiation --- in the most extreme case, no atmosphere may exist at all.  One may also study the effects of atmospheric chemistry, cloud cover, oceans and a hydrological cycle on the atmospheric circulation of exo-Earths \citep{j97,j03,p11}, but in the absence of observational constraints on the atmosphere of Gliese 581g we have chosen to omit these additional details.  As such observational constraints become available, the models should evolve in sophistication to be commensurate with the astronomical data.  When astronomical instrumentation becomes advanced enough to measure the phase curves from exo-Earths, then the relative importance of advection versus radiative cooling may be quantified \citep{ca11} and hence a subset of our models may be verified or refuted.

\section*{Acknowledgments}

K.H. acknowledges support from the Zwicky Prize Fellowship and the use of the \texttt{Brutus} computing cluster (adroitly managed by Olivier Bryde et al. ) at ETH Z\"{u}rich, as well as useful comments from Dick McCray and Rory Barnes on earlier versions of the manuscript.  S.V. acknowledges support from NSF grant AST-0307493.  K.H. is indebted to Kristen Menou for introducing him to this field of research and for many illuminating discussions.  We thank the anonymous referee for useful comments which improved the quality of the manuscript.
%\vspace{-0.2in}

%%% REFERENCES %%%

\label{lastpage}


\begin{thebibliography}{99}

\bibitem[Anderson et al.(2004)]{anderson04} Anderson, J.L., et al. \ 2004, Journal of Climate, 17, 4641

\bibitem[Charbonneau et al.(2009)]{char09} Charbonneau, D., et al. \ 2009, Nature, 462, 891

\bibitem[Cho et al.(2003)]{cho03} Cho, J.Y.-K., Menou, K., Hansen, B.M.S., \& Seager, S. \ 2003, ApJ, 587, L117

\bibitem[Cho et al.(2008)]{cho08} Cho, J.Y.-K., Menou, K., Hansen, B.M.S., \& Seager, S. \ 2008, ApJ, 675, 817

\bibitem[Cooper \& Showman(2005)]{cs05} Cooper, C.S., \& Showman, A.P. \ 2005, ApJ, 629, L45

\bibitem[Cooper \& Showman(2006)]{cs06} Cooper, C.S., \& Showman, A.P. \ 2006, ApJ, 649, 1048

\bibitem[Cowan \& Agol(2011)]{ca11} Cowan, N.B., \& Agol, E. \ 2011, ApJ, in press (arXiv:1011.0428v1)

\bibitem[Dvorak et al.(2010)]{dvo10} Dvorak, R., et al. \ 2010, Astrobiology, 10, 33

\bibitem[Edwards et al.(2004a)]{ed04} Edwards, H.G.M., Cockell, C.S., Newton, E.M., \& Wynn-Williams, D.D. \ 2004, Journal of Raman Spectroscopy, 35, 463

\bibitem[Edwards et al.(2004b)]{ed04b} Edwards, H.G.M., de Oliveira, L.F.C., Cockell, C.S., Ellis-Evans, J.C., \& Wynn-Williams, D.D. \ 2004, International Journal of Astrobiology, 3, 125

\bibitem[Fraser \& Brown(2010)]{fb10} Fraser, W.C., \& Brown, M.E. \ 2010, ApJ, 714, 1547

\bibitem[Gold(1992)]{gold92} Gold, T. \ 1992, Proceedings of the National Academy of Sciences, 89, 6045

\bibitem[Gordon \& Stern(1982)]{gs82} Gordon, C.T., \& Stern, W.F. \ 1982, Monthly Weather Review, 110, 625

\bibitem[Held \& Suarez(1994)]{hs94} Held, I.M., \& Suarez, M.J. \ 1994, Bulletin of the American Meteorological Society, 75, 1825

\bibitem[Heng, Menou \& Phillipps(2011)]{hmp11} Heng, K., Menou, K., \& Phillipps, P.J. \ 2011, MNRAS, in press (arXiv:1010.1257v3)

\bibitem[Joshi, Haberle \& Reynolds(1997)]{j97} Joshi, M.M., Haberle, R.M., \& Reynolds, R.T. \ 1997, Icarus, 129, 450

\bibitem[Joshi(2003)]{j03} Joshi, M. \ 2003, Astrobiology, 3, 415

\bibitem[Kraucunas \& Hartmann(2005)]{kh05} Kraucunas, I., \& Hartmann, D.L. \ 2005, Journal of Atmospheric Sciences, 62, 371

\bibitem[Lammer et al.(2010)]{lammer10} Lammer, H., et al. \ 2010, Astrobiology, 10, 45

\bibitem[Mayor et al.(2009)]{mayor09} Mayor, M., et al. \ 2009, A\&A, 507, 487

\bibitem[Menou \& Rauscher(2009)]{mr09} Menou, K., \& Rauscher, E. \ 2009, ApJ, 700, 887

\bibitem[Merlis \& Schneider(2010)]{ms10} Merlis, T.M., \& Schneider, T. \ 2010, Journal of Advances in Modeling Earth Systems -- Discussion (JAMES-D), 2, 13-1

\bibitem[Mitchell \& Vallis (2010)]{mv10} Mitchell, J.L., \& Vallis, G.K. \ 2010, Journal of Geophysical Research, in press (arXiv:1008.1996v1)

\bibitem[Peix\'{o}to \& Oort(1984)]{po84} Peix\'{o}to, J.P., \& Oort, A.H. \ 1984, Reviews of Modern Physics, 56, 365

\bibitem[Pierrehumbert(2011)]{p11} Pierrehumbert, R.T. \ 2011, ApJ, 726, L8

\bibitem[Roeckner \& von Storch(1980)]{rvs80} Roeckner, E., \& von Storch, H. \ 1980, Atmosphere-Ocean, 18, 239

\bibitem[Saravanan(1993)]{sar93} Saravanan, R. \ 1993, Journal of Atmospheric Sciences, 50, 1211

\bibitem[Scalo et al.(2007)]{scalo07} Scalo, J., et al. \ 2007, Astrobiology, 7, 85

\bibitem[Selsis et al.(2007)]{sel07} Selsis, F., Kastings, J.F., Levrard, B., Paillet, J., Ribas, I., \& Delfosse, X. \ 2007, A\&A, 476, 1373

\bibitem[Shapiro(1970)]{shapiro70} Shapiro, R. \ 1970, Reviews of Geophysics and Space Physics, 8, 359

\bibitem[Shell \& Held(2004)]{sh04} Shell, K.M., \& Held, I.M. \ 2004, Journal of the Atmospheric Sciences, 61, 2928

\bibitem[Showman, Cho \& Menou(2010)]{showman10} Showman, A.P., Cho, J.Y.-K., \& Menou, K. \ 2010, Exoplanets, ed. S. Seager, pg. 471--516 (Tucson: University of Arizona Press)

\bibitem[Showman \& Polvani(2010)]{sp10} Showman, A.P., \& Polvani, L.M. \ 2010, Geophysical Research Letters, 37, L18811

\bibitem[Smagorinsky(1963)]{sma63} Smagorinsky, J. \ 1963, Monthly Weather Review, 91, 99

\bibitem[Smagorinsky(1964)]{sma64} Smagorinsky, J. \ 1964, Quarterly Journal of the Royal Meteorological Society, 90, 1

\bibitem[Snellen et al.(2010)]{snellen10} Snellen, I.A.G., de Kok, R.J., de Mooij, E.J.W., \& Albrecht, S. \ 2010, Nature, 465, 1049

\bibitem[Stauffer \& Hartmann(1986)]{sh86} Stauffer, J.R., \& Hartmann, L.W. \ 1986, ApJ, 61, 531

\bibitem[Stephenson(1994)]{s94} Stephenson, D.B. \ 1994, Q.J.R. Meteorol. Soc., 120, 699

\bibitem[Suarez \& Duffy(1992)]{sd92} Suarez, M.J., \& Duffy, D.G. \ 1992, Journal of Atmospheric Sciences, 49, 1541

\bibitem[Tarter et al.(2007)]{tarter07} Tarter, J.C., et al., 2007, Astrobiology, 7, 30

\bibitem[Thrastarson \& Cho(2010)]{tc10} Thrastarson, H.Th., \& Cho, J.Y.-K. \ 2010, ApJ, 716, 144

\bibitem[Vallis(2006)]{vallis06} Vallis, G.K. \ 2006, Atmospheric and Oceanic Fluid Dynamics: Fundamentals and Large-Scale Circulation (New York: Cambridge University Press)

\bibitem[Vogt et al.(2010)]{vogt10} Vogt, S.S., Butler, R.P., Rivera, E.J., Haghighipour, N., Henry, G.W., \& Williamson, M.H. \ 2010, ApJ, 723, 954

\bibitem[Washington \& Parkinson(2005)]{wp05} Washington, W.M., \& Parkinson, C.L. \ 2005, An Introduction to Three-Dimensional Climate Modeling, second edition (Sausalito: University Science Books)

\bibitem[Watkins \& Cho(2010)]{wc10} Watkins, C., \& Cho, J.Y.-K. \ 2010, ApJ, 714, 904

\bibitem[Wordsworth et al.(2011)]{word10} Wordsworth, R.D., Forget, F., Selsis, F., Madeleine, J.-B., Millour, E., \& Eymet, V. \ 2011, A\&A, in press (arXiv:1005.5098v2)

\bibitem[Wordsworth et al.(2010)]{word10b} Wordsworth, R.D., Forget, F., Millour, E., Madeleine, J.-B., Eymet, V., \& Selsis, F. \ 2010, ASP Conference Series, ``Pathways Towards Habitable Planets", 430, 558

\end{thebibliography}
\end{document}